\documentclass[superscriptaddress,amsmath,amssymb,pre,twocolumn,floatfix]{revtex4-1}
\usepackage{graphicx,subfigure}
\usepackage{dcolumn}
\usepackage{bm}
\usepackage{marginnote}
\usepackage{wrapfig}
\usepackage{color}
\usepackage{epsf,epsfig}
\usepackage{color}
\usepackage{enumerate}
\usepackage[colorlinks=true,allcolors=blue]{hyperref}
\newcommand{\kt}[0]{\tilde{K}_r(r)}
\newcommand{\ktone}[0]{\tilde{K}_r^\prime(r)}
\newcommand{\kttwo}[0]{\tilde{K}_r^{\prime\prime}(r)}
\newcommand{\ktthree}[0]{\tilde{K}_r^{\prime\prime\prime}(r)}
\newcommand{\rt}[0]{\tilde{r}}

\newcommand{\pr}[0]{^\prime}
\newcommand{\ppr}[0]{^{\prime\prime}}
\newcommand{\va}[0]{(r,t)}
\newcommand{\var}[0]{(r,t(r))}
\newcommand{\co}[0]{\text{cos}}

\begin{document}
\title{Buckling without bending morphogenesis: \\Nonlinearities, spatial confinement, and branching hierarchies}
\author{M. C. Gandikota}
\email{mgandiko@syr.edu}
\affiliation{Physics Department and BioInspired Institute, Syracuse University, Syracuse, NY 13244 USA}
\author{J. M. Schwarz}
\email{jschwarz@physics.syr.edu}
\affiliation{Physics Department and BioInspired Institute, Syracuse University, Syracuse, NY 13244 USA}
\affiliation{Indian Creek Farm, Ithaca, NY 14850 USA}
\date{June 29, 2021}

\begin{abstract}
During morphogenesis, a featureless convex cerebellum develops folds. As it does so, the cortex thickness is thinnest at the crest (gyri) and thickest at the trough (sulci) of the folds. This observation cannot be simply explained by elastic theories of buckling. A recent minimal model explained this phenomenon by modeling the developing cortex as a growing fluid under the constraints of radially spanning elastic fibers, a plia membrane and a nongrowing sub-cortex (Engstrom, {\it et. al., PRX} 2019). In this minimal buckling without bending morphogenesis (BWBM) model, the elastic fibers were assumed to act linearly with strain.  Here, we explore how nonlinear elasticity influences shape development within BWBM.  The nonlinear elasticity generates a quadratic nonlinearity in the differential equation governing the system's shape and leads to sharper troughs and wider crests, which is an  identifying characteristic of cerebellar folds at later stages in development. As developing organs are typically not in isolation, we also explore the effects of steric confinement, and observe flattening of the crests.  Finally, as a paradigmatic example, we propose a hierarchical version of BWBM from which a novel mechanism of branching morphogenesis naturally emerges to qualitatively predict later stages of the morphology of the developing cerebellum.  
\end{abstract}

\maketitle
\flushbottom
\section{Introduction}\label{introduction}

When morphogenesis is viewed through the lens of a physicist, one of the typical mechanistic routes to take is morphoelasticity induced by varying internal stresses~\cite{ben_amar_2019}. Shape change in response to internal stresses, within elastic objects which tend to retain shape, provides us with a large collection of shapes. Differential growth can be a source of such an internal stress. This reasonable viewpoint is made manifest in the widespread use of the Euler buckling instability in purely elastic materials to explain the onset of folds in morphogenesis problems~\cite{nelson2016buckling} as diverse as cerebra \cite{budday2015physical,richman1975mechanical,raghavan1997continuum,bayly2013cortical,tallinen,tallinen2016growth}, intestinal crypts and villi \cite{hannezo2011instabilities,shyer2013villification}, airway mucus wrinkles \cite{wiggs1997mechanism,li2011surface}, tooth ridges \cite{osborn2008model} and hair-follicle patterns \cite{shyer2017emergent}. Recent work accounting for cerebellar foliation falls under the same framework ~\cite{lejeune,lejeune2019}. These models typically predict a characteristic length scale between folds and state quantitative agreement between prediction and observation to validate such an approach.

On the other hand, tissue fluidity in early stages of developing embryos has emerged as a driver of shape change in both zebrafish and the insect {\it Tribolium castaneum}~\cite{mongera_2018,jain_2020}. Fluids, unlike elastic solids, do not retain their shape. Then how does fluidity - the antithesis of elasticity - affect the shape of the developing organ? Presumably biological systems have figured out ways to combine elements of elasticity and fluidity to bring about an even more complex collection of shapes at later stages of development. The buckling without bending morphogenesis (BWBM) model is one such model affirming the cleverness of biological systems~\cite{engstrom2018}. Its origin is rooted in explaining recent quantitative observations of the developing murine cerebellum, a much less studied component of the brain as compared to the cerebrum. 

Vertebrate brains are typically divided into three different sectors -- the forebrain, the midbrain, and the hindbrain. While the forebrain consists of the cerebrum, the hypothalamus, and the thalamus, the hindbrain contains the spinal cord, medulla oblongata, and the cerebellum.  Since the shape of the forebrain varies more across species than the midbrain and hindbrain, its development has been the focus of much study.  On the other hand, by studying conserved patterns in the hindbrain, we gain complementary insight into the morphogenetic processes of the brain. 

To be specific, let us characterize the difference in shape between the cerebrum and cerebellum.  First, the shape of the cerebellum has an approximate cylindrical symmetry, whereas such a symmetry is absent in the cerebrum. Second, mammalian cerebellums of all sizes produce folds~\cite{larsell}, whereas small mammalian cerebrums do not show folds~\cite{herculano-houzel_2009}. Given that the overall size of \textit{both} organs increases with increasing species size, it is possible that the differences in shape arises via different physical shape change mechanisms. If so, do different physical shape change mechanisms lead to different emergent functionalities of the two? Moreover, the conservation of the number of the 8-10 primary lobes of the cerebellum across species demands a treatment on its own footing as it suggests a scale invariant shape change mechanism. Number of secondary folds however, can differ across species. 

Recent experimental observations of the developing murine cerebellum found that the proliferating cells in the cerebellar cortex are motile with neighbor exchanges on the order of minutes~\cite{lawton2019}. It was also observed that the developing cerebellar cortex varies in thickness with the trend being that the cortex is thinnest at the crest (gyri) and thickest at the trough (sulci). Moreover, the developing cerebellum is under \textit{tension} and not under compression as evidenced by both radial and circumferential cuts. While these observations cannot be explained by elastic wrinkling theories, all three of these findings can be explained by the linear BWBM model in which a growing cerebellar cortex is \textit{fluid-like} and the sub-cortex is a nongrowing core~\cite{engstrom2018}. The cortex is under tension due to the presence of radial glial cells spanning the cerebellum as well as the pial membrane. Finally, Bergmann glial cells spanning the cortex attempt to maintain constant thickness of the cortex. See Fig. \ref{schematic}. Cell growth in the presence of such constraints drives a featureless convex cortex to form folds. In addition, the BWBM model also offers an explanation for the length scale invariance of the formation of cerebellum folds to understand the conservation of 8-10 primary lobes across vertebrate species spanning a range of sizes~\cite{engstrom2018}. 

\begin{figure}[tbh]
\includegraphics[width=0.45\textwidth]{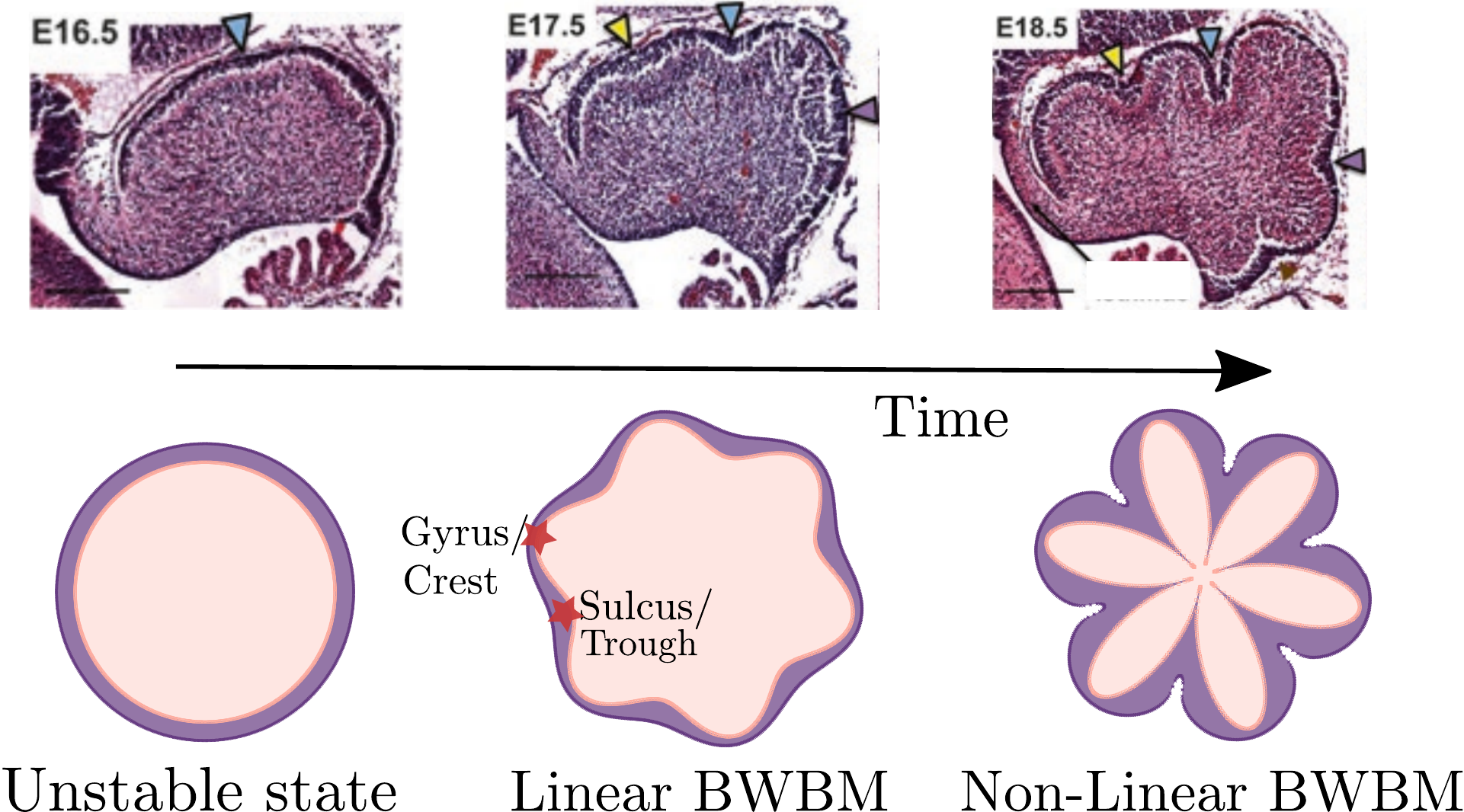}
\caption{{\it Visual comparison of experiment and model.} The morphogenesis of cerebellum involves the development of folds in an initially featureless convex cerebellum. The top row shows midsagittal cross-sections of mouse cerebellum prior to birth. Scale bar is 200 $\mu$m. The arrowheads are from the original figures in Ref. \cite{marzban} and indicate fissures.  Reprinted with permission from Springer Nature. The linear BWBM model reproduces the smooth sinusoidal like folds that occur at the onset of folding. The BWBM model with nonharmonic radial glial springs reproduces the sharp cusped sulci that we see at the later stages of cerebellar development. Note that time is not explicit in the quasi-static BWBM. We start from system parameters describing the unstable state and minimize the energy while obeying constraints to arrive at the nonlinear BWBM configuration. We use brain folding terminology - gyri (sulci) and oscillator terminology - crests (troughs) interchangeably to refer to hills (valleys) of oscillations respectively. \textit{Parameters:} $c=0.1,0.155,\epsilon=0.6,\rho=31.3,\tilde{t}_0=0.1,0.37,\tilde{k}_{r_1}=0,-0.97,\tilde{k}_{r_2}=0,1.05$ for the linear and nonlinear BWBM respectively. [$\tilde{r},d\tilde{r}/d\theta]_{\theta=0} \mbox{ is } [2.01,-0.83]$.}\label{expt comparison}
\end{figure}

The linear BWBM model provides a quantitative framework for the onset of shape change that manifest as smooth cortex oscillations in the developing cerebellum. However, as the shape of the cerebellum continues to evolve, we observe cusped sulci and wide gyri (see Fig. \ref{expt comparison}). The linear BWBM model, which can be mapped to a forced, simple harmonic oscillator, cannot account for such nonlinear phenomena. These observations necessitate the exploration of nonlinear elasticity within the BWBM model, particularly in the context of tensioned radial glial cells given that robust nonlinearities have been observed in stretched cells~\cite{fernandez_2006}. In addition to exploring the effect of nonlinear springs in the BWBM model, we will also explore the impact of spatial confinement on shape change, which imposes a different form of nonlinearity in the BWBM model. Finally, since we are exploring stages of shape development beyond the onset of shape change, we note the hierarchy of folds that emerges in the developing cerebellum. This hierarchy manifests itself as folds within folds. As the developing cerebellum grows in size, this leads to a type of branching morphogenesis. While branching morphogenesis has been thought of in the context of developing lungs, kidneys, and other organs~\cite{varner_2014,hannezo_2017}, a hierarchical version of BWBM, given its scale invariant solutions, naturally emerges as a potential candidate to explain branching morphogenesis within the cerebellum.  

The outline of the manuscript is as follows. In Section \ref{linear}, we review the linear BWBM model focusing on the early stages of shape change in the developing cerebellum. Section \ref{nonlinear_kr} discusses the effect of nonlinear springs that describe the radial fibrous glial cells and its role in sharpening the sinusoidal sulci of the linear model. Section \ref{tanh_constraint} discusses the effect of steric hindrances on the flattening of gyri and Section \ref{hierarchy} discusses the hierarchy of subsequent folds emerging at even later stages of cerebellar development. Section \ref{discussion} summarizes the work and addresses its implications.

\section{Linear buckling without bending morphogenesis model}\label{linear}
\begin{figure}[h!]
\includegraphics[width=0.48\textwidth]{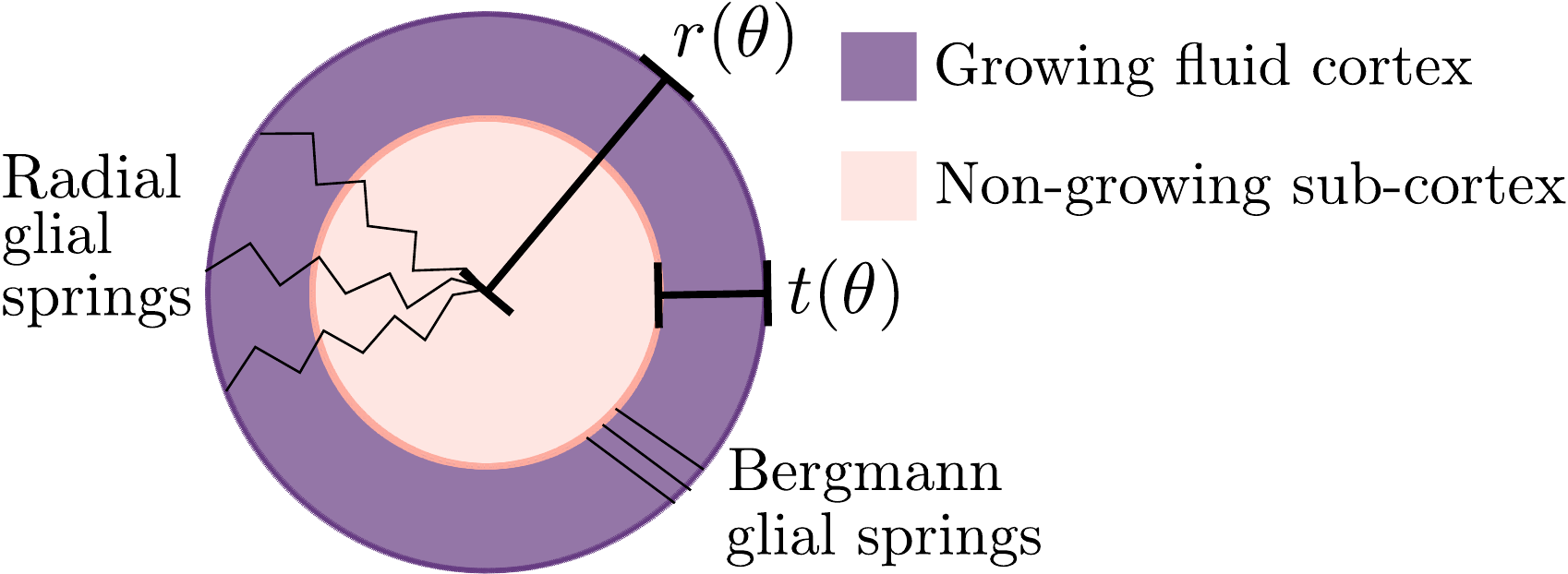}
\caption{{\it Schematic of the buckling without bending morphogenesis (BWBM) model.} The radial glial cells span the cerebellum and Bergmann glial cells span the cortex and resist thickness variations in the cortex. The area of the sub-cortex does not grow as fast as the cortex and so it is approximated as nongrowing over some time scale.}\label{schematic}
\end{figure}
Within a timespan of a day, the featureless, convex developing cerebellum begins to develop folds. The BWBM model provides a physical basis for the onset of these folds.  Unlike the cerebrum, the cylindrical symmetry of the cerebellum allows for two-dimensional modeling of its sagittal cross section. The two-dimensional BWBM model \cite{engstrom2018} offers a quasi-static description of the foliation i.e., it does not describe the dynamics of growth but determines the final equilibrium shape for a given set of parameters. As the parameters change with time, so does the shape. In polar coordinates, the radius of the cerebellum and the thickness of the cortex, otherwise known as the external granular layer (EGL), are represented with $r,t$ respectively (see Fig. \ref{schematic}). The energy functional for the bi-layer model with $\theta$ parametrizing the two degrees of freedom is,
\begin{small}
\begin{equation}\label{energy_functional}
E\left[r,t,\frac{dt}{d\theta}\right]=\int d\theta\left\{k_{r_0}(r-r_0)^2-k_t(t-t_0)^2+\beta\left(\frac{dt}{d\theta}\right)^2\right\}.
\end{equation}
\end{small}

\noindent Here, all the constants $k_{r_0},k_t,\beta,r_0,t_0$ are positive. The first term represents the elastic contribution of the radial glial fibrous cells spanning the cerebellum and the elastic pial membrane surrounding the cerebellum. Moreover, $k_r$ is the elasic modulus for this term with $r_0$ as its rest length radius. The second term is a growth potential for the cortex with $k_t$ controlling the contribution of the term and $t_0$ setting the rest length thickness. The third term represents the Bergmann glial fibrous cells resisting thickness variations of the cortex with $\beta$ being the accompanying elastic modulus. Note that the third term is \textit{not} a bending energy term. A bending energy term would involve the curvature as given by the second derivative with $\theta$, unlike the first derivative used here. Finally, given the slower growth of the sub-cortex/core, as compared to the cortex, we demand that the area of the sub-cortex does not change, at least over the time scale of the onset of the foliation, i.e. a day.  In other words, 
\begin{equation}\label{subcortex_area}
\frac{1}{2}\int d\theta(r-t)^2=A_0=\text{constant}.
\end{equation}
It is the area conservation that sets up the competition between the radial elastic energy of the glial fibers and pial membrane with the growth potential term. 

\noindent The extremum of the variational problem subject to this constraint yields a pair of coupled Euler-Lagrange equations for $r,t$. The solution to these equations are linear sinusoidal functions of the form,

\begin{equation}
\begin{split}
t(\theta)&=T\;\text{sin}(n\theta+\phi),\\
r(\theta)&=\left(\frac{1}{1-\epsilon}\right)r_0(\theta)-\left(\frac{\epsilon}{1-\epsilon}\right)t(\theta),
\end{split}
\end{equation} 
where 
\begin{flalign*}
n&=\sqrt{\rho(1+\epsilon c/(1-\epsilon))},\\
T&=\sqrt{2}\epsilon\sqrt{\frac{A_0}{\pi}-\left(\frac{a-t_0}{1-\epsilon+\epsilon c}\right)^2}. 
\end{flalign*}
The dimensionless constants are $c:=k_{r_0}/k_t,\epsilon:=\mu/k_{r_0}$ and $\rho:=k_t/\beta$.

The linear BWBM model successfully explains the cortex thickness oscillations being out of phase with the sub-cortex deformation for a developing cerebellum which leads to a thinner cortex at the crest and a thicker cortex at the trough. This model also allows the amplitude of the cortex thickness oscillation to be the same size or even greater than the amplitude of the substrate oscillations. Such phenomena is in contrast to elastic wrinkling models which fail to wrinkle for thicker cortices at the trough~\cite{lagrange2016wrinkling}. 

\section{nonlinear elasticity}\label{nonlinear_kr}
While the linear BWBM model addresses the onset of shape change, a next follow up question to ask is -- what are the limitations of the linear BWBM model in explaining the more dramatic shape changes in the developing cerebellum at later stages of development?  As the radial glial cells and the pial membrane become more stretched by the developing crests, the enhanced stretching may lead to detachment of the radial glial cells from the pial membrane or may lead to nonlinear elastic effects. It has long been known that stretched cells act as nonlinear springs~\cite{fernandez_2006} and that collagen, which is one of dominant fibrous proteins constituting the pial membrane exhibits nonlinear elasticity as strain increases~\cite{storm_2005}. Such effects are not addressed in the linear BWBM model. It would therefore be prudent to examine the role of nonlinear elasticity in the BWBM model. To do so, we promote the radial spring constant $k_{r_0}$ to $k_r(r)$, a radially dependent spring `constant' -
\begin{equation}
k_r(r):=k_{r_0}+k_{r_1}(r-r_0)+k_{r_2}(r-r_0)^2.
\end{equation}
The above three terms correspond to quadratic, cubic and quartic energy terms of the radial glial spring potential energy. The cubic energy term brings an asymmetry in the radial spring energy across $r_0$ and the quartic energy term counteracts the destabilizing effect of the cubic term. The more general form of the uncoupled differential equation where every spring constant is allowed to be nonlinear is explored in Appendix \ref{general uncoupled}. We nondimensionalize the problem as $\tilde{E}:=E/(k_{r_0}r_0^2), \tilde{r}:=r/r_0,\tilde{t}:=t/r_0,\tilde{t}_0:=t_0/r_0$ and

\begin{equation}
\kt:=\frac{k_r(r)}{k_{r_0}}=1+\tilde{k}_{r_1}(\rt-1)+\tilde{k}_{r_2}(\rt-1)^2,
\end{equation}

\noindent where $\tilde{k}_{r_1}=k_{r_1}r_0/k_{r_0}$ and $\tilde{k}_{r_2}=k_{r_2}r_0^2/k_{r_0}$. The nondimensional energy functional with dimensionless variables and coefficients is,
\begin{equation}\label{functional_nonlinear_kr}
\begin{split}
\tilde{E}\left[\tilde{r},\tilde{t},\frac{d\tilde{t}}{d\theta}\right] = \int d\theta\bigg\{&\kt(\tilde{r}-1)^2-\frac{1}{c}(\tilde{t}-\tilde{t}_0)^2 \\
&+ \frac{1}{\rho c}\left(\frac{d\tilde{t}}{d\theta}\right)^2\bigg\},
\end{split}
\end{equation}
whose minimization is subject to the constraint, 
\begin{equation}
\frac{1}{2}\int d\theta(\tilde{r}-\tilde{t})^2=\frac{A_0}{r_0^2}=\text{dimensionless constant}.
\end{equation}

\noindent The variational problem at hand is,
\begin{equation}\label{lagrange}
\delta\left[\tilde{E}-\epsilon\int d\theta(\tilde{r}-\tilde{t})^2\right]=0
\end{equation}

\noindent The corresponding Euler-Lagrange equations are 
\begin{equation}\label{r_genkr}
\kt (\rt-1) - \epsilon(\rt-\tilde{t})+\frac{\ktone}{2}(\rt-1)^2=0,
\end{equation}

\begin{equation}\label{genkr1}
\frac{1}{\rho c}\frac{d^2\tilde{t}}{d\theta^2}+\frac{1}{c}(\tilde{t}-\tilde{t}_0)-\epsilon(\tilde{r}-\tilde{t})=0.
\end{equation}
For the purpose of numerically solving these coupled differential equations, we may as well stop here. However, the exercise of uncoupling the differential equations points to an important source of nonlinearity. Towards finding it, we differentiate Eq. \ref{r_genkr} twice with $\theta$ to arrive at
\begin{equation}\label{genkr2}
\begin{split}
&\left[\kt+2(\rt-1)\ktone+\frac{(\rt-1)^2}{2}\ktthree-\epsilon\right]\left(\frac{d^2\rt}{d\theta^2}\right)\\
+&\left[3\ktone+3(\rt-1)\kttwo+\frac{(\rt-1)^2}{2}\ktthree\right]\left(\frac{d\rt}{d\theta}\right)^2\\
=&-\epsilon\frac{d^2\tilde{t}}{d\theta^2}.
\end{split}
\end{equation}

\noindent Here we see the nonlinear term $(d\tilde{r}/d\theta)^2$. This is a  consequence of the inhomogeneous radial spring constant and the coupling between the Euler-Lagrange equations brought about by the Lagrange multiplier in Eq. \ref{lagrange}. Solving Eq. \ref{r_genkr} for $\tilde{t}$, we have,
\begin{equation}\label{genkr4}
\tilde{t}=\rt-\frac{1}{\epsilon}\left[(\rt-1)\kt+\frac{\ktone}{2}(\rt-1)^2\right].
\end{equation}
Substituting Eq. \ref{genkr4} in Eq. \ref{genkr1} leads to,
\begin{equation}\label{genkr5}
\begin{split}
\frac{d^2\tilde{t}}{d\theta^2}=\rho\bigg[&\bigg(\bigg\{(\rt-1)\kt+\frac{\ktone}{2}(\rt-1)^2\bigg\}-\rt\bigg)\\
\times&\bigg(\frac{1+\epsilon c}{\epsilon}\bigg)+\epsilon c\rt +\tilde{t}_0\bigg].
\end{split}
\end{equation}

\noindent Substituting, Eq. \ref{genkr5} in Eq. \ref{genkr2}, we find the shape equation of the system to be,
\begin{equation}\label{general_kr_EL}
\begin{split}
&~~~\left[\kt+2(\rt-1)\ktone + \frac{(\rt-1)^2}{2}\kttwo-\epsilon \right]\left(\frac{d^2\tilde{r}}{d\theta^2}\right)\\
&+\left[3\ktone+3(\rt-1)\kttwo+\frac{(\rt-1)^2}{2}\ktthree\right]\left(\frac{d\tilde{r}}{d\theta}\right)^2\\
&-\rho\epsilon \rt \\
&+\rho(1+\epsilon c)\left[(\rt-1)\kt+\frac{(\rt-1)^2}{2}\ktone\right]\\
&+\rho\epsilon t_0=0.
\end{split}
\end{equation}
The importance of the term $(d\tilde{r}/d\theta)^2$, a source of nonlinearity in Eq. \ref{general_kr_EL}, lies in the robustness of its appearance. Irrespective of the form of nonlinearity in $k_r(r)$, we retain this nonlinear term whereas the coefficients in Eq. \ref{general_kr_EL} correspondingly vary (see Appendix \ref{general uncoupled}). Such a nonlinearity is also seen when one attempts to uncouple the Lotka-Volterra equations \cite{davis1961}. 

We use the RK45 method of \texttt{scipy.integrate} package in python for numerical integration of all differential equations in this paper. For generating the phase-portraits in Fig. \ref{phase portraits panel}, we use \texttt{XPPAUT} \cite{ermentrout}.  

\begin{figure*}[tbh]
\includegraphics[width=\textwidth]{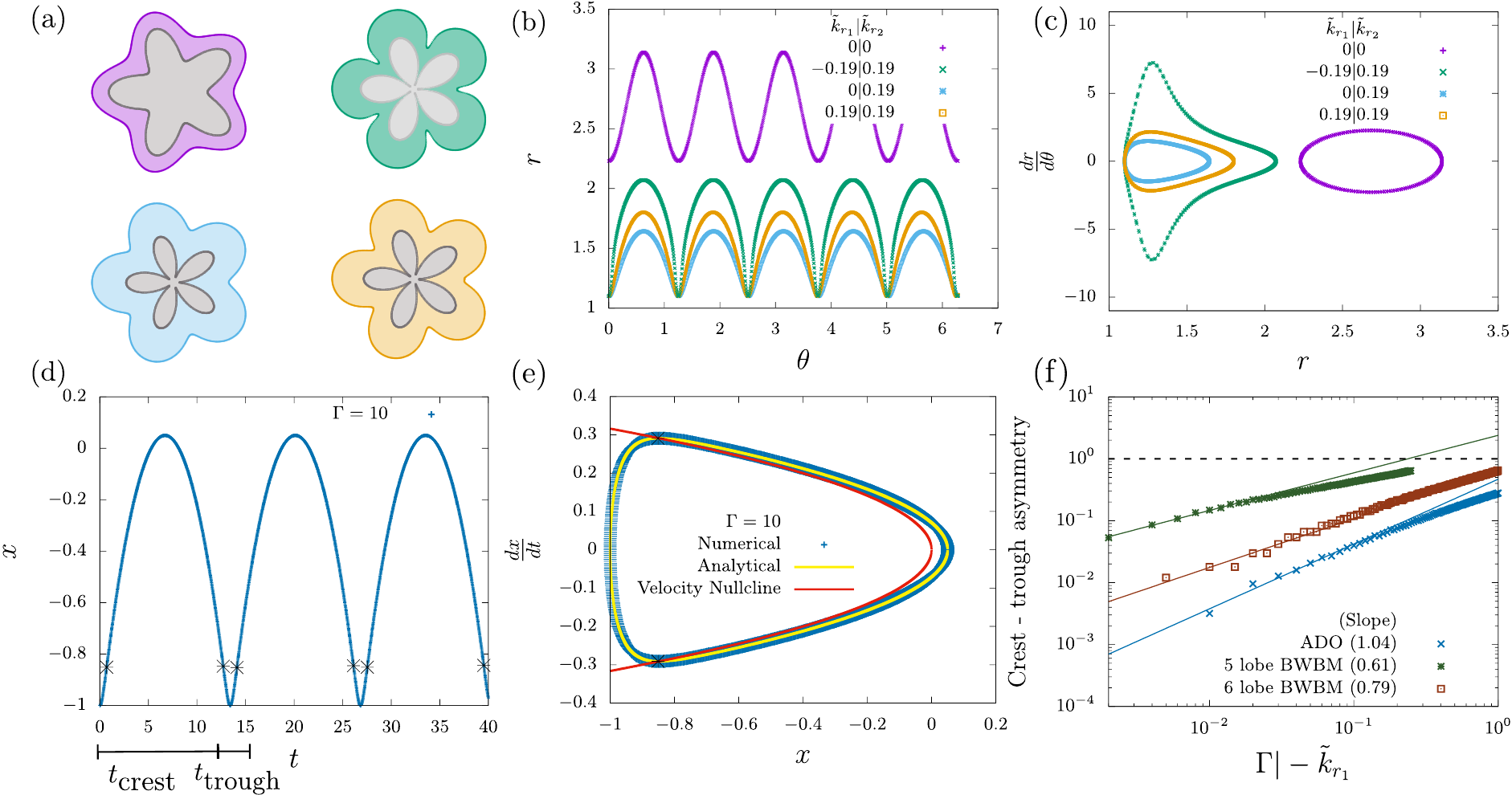}
\caption{\textit{Asymmetry in the widths of gyri (crests) and sulci (troughs).} a,b) Numerical solutions in polar and Cartesian coordinates for BWBM with nonharmonic radial glial springs. Nonsinusoidal sharp sulci can be seen for systems with nonzero $\tilde{k}_{r_1},\tilde{k}_{r_2}$ (green, blue, yellow). c) Phase space orbits of solutions in (a,b). The left-right asymmetry in the orbits seen as a steeper drop on the left translates to sharper troughs in the coordinate space. d) Numerical solution in Cartesian coordinates for the ADO oscillator. Sharper troughs are observed due to the addition of the nonlinear force term $(dx/dt)^2$ to the simple harmonic oscillator. The acceleration of the oscillator is zero at the points marked with the asterisk. The width of crest and trough is shown. e) Numerical and analytical solution of the phase-space orbit for the ADO. The velocity nullcline, in red, is the locus of points with zero acceleration. The position coordinate (marked in asterisk) at which the nullcline intersects the orbit decides the widths of crests and troughs in (d). The shape of the orbit is seen to be especially similar to the orbits in (c) for the cases of $\tilde{k}_{r_1}=0,0.19$. f) Crest - trough asymmetry for the ADO and the BWBM model as a function of their respective tuning parameters: $\Gamma$ and $-\tilde{k}_{r_1}=\tilde{k}_{r_2}$. The dashed black line is the upper limit of the asymmetry measure. \textit{Parameters} 5-lobe BWBM in (a)-(c) and (f): $c=0.1809,\epsilon=0.9,\rho=15.6,\tilde{t}_0=0.7$. For periodicity in (a), parameter $c$ is appropriately tuned for choices of $\tilde{k}_{r_{1,2}}$. 6-lobe BWBM in (f): $c=0.1,\epsilon=0.6,\rho=31.3,\tilde{t}_0=0.5$. $[\tilde{r},d\tilde{r}/d\theta]_{\theta=0}$ is [1.1,0],[1.25,0] for 5-/6- lobe.}\label{asymmetry panel}
\end{figure*}

Results of numerical integration in Fig. \ref{asymmetry panel}a,b, show asymmetric oscillations - sharper sulci and smooth gyri. This holds true even with $\tilde{k}_{r_1}=0$, i.e. with no explicit imposed asymmetry in the energy functional (see Eq. \ref{functional_nonlinear_kr}). This points to the possible role of the $(d\tilde{r}/d\theta)^2$ nonlinearity in bringing about asymmetric oscillations. In the context of explaining the sharp sulci in the cerebral cortex of larger mammals, the sulcification instability in nonlinear elastic materials has been used \cite{tallinen}. Here we observe, sans an elastic instability, sharper sulci in comparison to their sinusoidal counterparts. Even under the influence of these nonlinearities, the system robustly retains the thicker cortical troughs and thinner cortical crests observed in the linear BWBM model.  For small $\tilde{k}_{r_1},\tilde{k}_{r_2}$, the number of folds do not change appreciably in comparison to the linear BWBM model. 

\subsection{Assisting-dampening oscillator}\label{ado}
In Sec. \ref{nonlinear_kr}, the nonharmonic radial glial springs generate a shape equation \ref{general_kr_EL} which has several sources of nonlinearity. Given the assured presence of the quadratic nonlinearity $(d\tilde{r}/d\theta)^2$ in the shape equation irrespective of the nonlinearity introduced, we seek to isolate the effect it has in determining the shape of the system. Towards that end, we study the simple harmonic oscillator with a velocity dependent force in this section.

The differential equation of motion for a one dimensional, simple harmonic oscillator of mass $m$ with an external nonlinear forcing term, $-\gamma(d\tilde{x}/dt)^2$ is written as, 
$$m\frac{d^2\tilde{x}}{d\tilde{t}^2} = -k\tilde{x} - \gamma \left(\frac{d\tilde{x}}{d\tilde{t}}\right)^2.$$ 
The quadratic term assists/dampens the negative/positive velocity respectively bringing an asymmetry in the problem. We refer to this unphysical oscillator as the `assisting-dampening oscillator' (ADO). The parameters in this equation and the length scale $x_0$ from the initial condition $\tilde{x}(0)=\tilde{x}_0$ can be used to introduce nondimensional variables, $x:=\tilde{x}/x_0$ and $t:= \tilde{t}/\sqrt{m/k}$. The equation of motion effectively determined by a single tuning parameter $\Gamma := \gamma \tilde{x}_0/m$ is, 
\begin{equation}\label{ADO}
\frac{d^2x}{dt^2} = -x - \Gamma \left(\frac{dx}{dt}\right)^2.
\end{equation}
The nonlinear forcing term is nonconservative since it cannot be written down as the gradient of a position dependent potential. This problem can also not be formulated within the framework of Lagrangian mechanics as this velocity dependent force can not be represented by a function $U(x,\dot{x})$ such that,
$$- \Gamma \dot{x}^2=-\frac{\partial U}{\partial x}+\frac{d}{dt}\left(\frac{\partial U}{\partial \dot{x}}\right),$$
where $\dot{x}=dx/dt$~\cite{goldstein}. However, the time reversal symmetry of Eq. \ref{ADO} ensures all trajectories sufficiently close to the equilibrium point, ($x_0,y_0)=(0,0)$ to be closed orbits in the phase space \cite{strogatz}. This translates to periodic motion in the position-time space (see Fig. \ref{phase portraits panel}). 

The second order differential equation \ref{ADO} can be represented as two coupled first order differential equations,
\begin{equation}
\begin{split}
\frac{dx}{dt}&=y,\\
\frac{dy}{dt}&=-x-\Gamma y^2.
\end{split}
\end{equation}
Dividing the set of equations, we have,
\begin{equation}
y\frac{dy}{dx}=-x-\Gamma y^2.
\end{equation}
In contrast to the differential equation of motion in position-time space (see Eq. \ref{ADO}), the phase-space orbit is represented by a first-order differential equation. This makes finding an exact solution to the orbit in phase-space simpler. The variable transformation, $u=y^2$ linearizes the above differential equation as,
\begin{equation}\label{linear_DHO}
\frac{du}{dx}+2\Gamma u=-2x.
\end{equation}
The solution to the homogeneous differential equation,
\begin{equation}
\frac{du}{dx}+2\Gamma u=0,
\end{equation}
is $u(x)=A\;e^{-2\Gamma x}$ where $A$ is a constant that needs to be fixed by the initial condition. The particular solution to Eq. \ref{linear_DHO} is $u(x)=-x/\Gamma +1/(2\Gamma^2)$. Reverting to $y$, the total solution satisfying the initial condition $y(-1)=0$ is,
\begin{equation}\label{orbit_ADO_exact}
y(x)=\frac{1}{\sqrt{2}\Gamma} \sqrt{1-2\Gamma x- (1+2\Gamma )e^{-2\Gamma (x+1)}}.
\end{equation}
In the limit of $\Gamma\rightarrow0$, we have $y(x)\sim\pm\sqrt{1-x^2}$ which are  semi-circular arcs as expected for a simple harmonic oscillator. 

The numerical integration of the differential equation is presented in Fig. \ref{asymmetry panel}. The sharp troughs in the position-time space are solely due to the effect of the quadratic nonlinear term $(dx/dt)^2$. 

The red lines in Figs. \ref{asymmetry panel}e, \ref{phase portraits panel} are velocity nullclines which are the locus of points in phase-space where the acceleration, $dy/dt=d^2x/dt^2=0$. From Eq. \ref{ADO}, it follows that the nullcline is a quadratic function in $x$, 
\begin{equation}\label{ADO_null}
x=-\Gamma\left(\frac{dx}{dt}\right)^2.
\end{equation}
It is interesting to note that the quadratic nullcline (see Fig. \ref{phase portraits panel}b) bears resemblance to a similarly curved nullcline of the BWBM with nonharmonic springs (see Fig. \ref{phase portraits panel}d). The nullcline's curvature is dictated by the same quadratic nonlinear term $(dx/dt)^2$ in both cases.

For comparison with this unphysical oscillator, we also study a conservative Hamiltonian system and observe that it is also capable of exhibiting oscillations with sharp troughs. The simplest such system is obtained by adding cubic and quartic potential energy terms to the simple harmonic oscillator. The former term provides an explicit asymmetry in the energy functional, and the latter term ensures stability about the equilibrium point. In Appendix \ref{cubic_quartic_HO}, this system is explored numerically and analytically using a perturbation series constructed via the Lindstedt-Poincar\'e method \cite{jordan2007}.  

\begin{figure*}[tbh]
\includegraphics[width=1\textwidth]{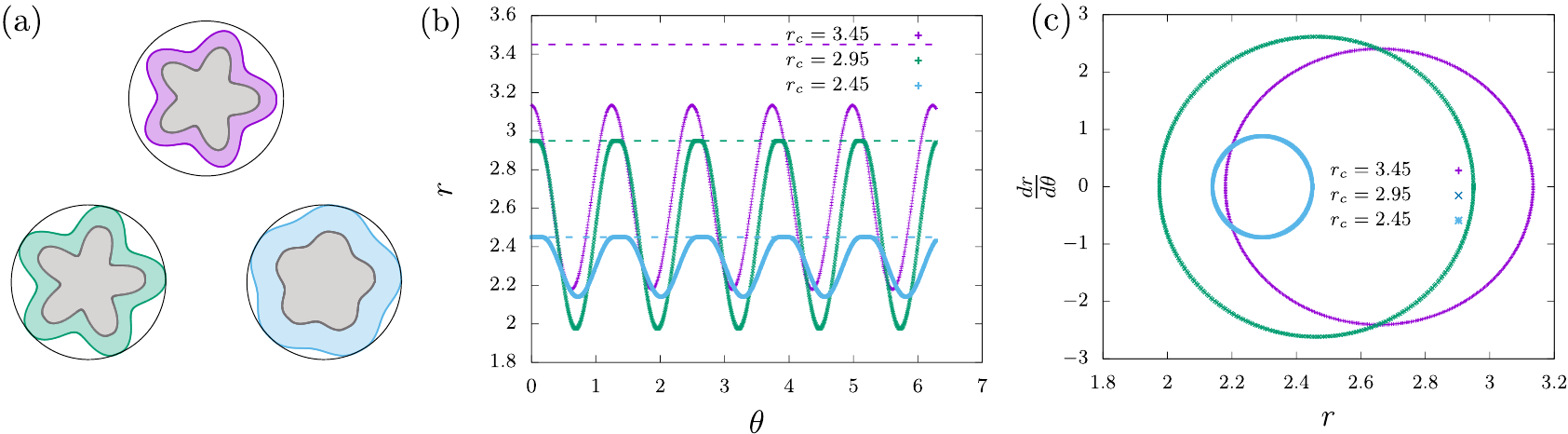}
\caption{\textit{Space constraint locally flattens the lobes of the linear BWBM model.} a,b) Polar and cartesian representation of the solutions to the differential equation of the linear BWBM model under a $\text{tanh}(\tilde{r}-\tilde{r}_c)$ space constraint (see Eq. \ref{nondim_tanh_circ}). The black confining circles in (a) are the confining walls and are represented by dashed lines in (b). c) Phase-space orbits, unlike the nonlinear BWBM model and the ADO model, remain symmetric. \textit{Parameters:} $c=0.07,\epsilon=0.9,\rho=15.6,\tilde{K}_c=30,q=10^4,\tilde{t}_0=7q,\tilde{r}_0=1q$. [$\tilde{r},\frac{d\tilde{r}}{d\theta}]_{\theta=0} \mbox{ is } [\tilde{r}_c,0]$.}\label{tanh panel}
\end{figure*}

\subsection{A measure for crest - trough asymmetry}
Stokes, in his study of propagating waves approaching the shore, discussed the development of narrow crests and wider troughs \cite{stokes}. In this context, measures for velocity and acceleration skewness have been proposed and studied. In the folds of cerebellum, which is a stationary spatial oscillation, a similar asymmetry presents itself in the form of wide gyri/crest and sharp sulci/trough. We propose the following measure for quantifying the asymmetry in widths -
\begin{equation}\label{cta}
\text{crest - trough asymmetry}:=\frac{t_{\text{crest}}-t_{\text{trough}}}{t_{\text{crest}}+t_{\text{trough}}}.
\end{equation}
Here, the length the parameter $\theta$ traverses between consecutive pair of points at which $d^2x/dt^2=0$ on the climbing (falling) wave,  and the immediately following falling (climbing) wave defines $t_{\text{crest}}$ ($t_{\text{trough}}$), respectively (see Fig. \ref{asymmetry panel}d). These points are the points of intersection of the phase-space orbit with the velocity nullcline (see Fig. \ref{asymmetry panel}e). The horizontal mouthed parabolic-shaped nullclines, thus influence the position of these points of intersection and play a prominent role in bringing the asymmetry between the widths of the crests and troughs. For the ADO, the velocity nullcline is \textit{exactly} parabolic (see Eq. \ref{ADO_null} and Fig. \ref{phase portraits panel}). The parabolic shape of the nullcline is due to the quadratic nonlinearity $(dx/dt)^2$ in Eq. \ref{ADO}. For the BWBM model with nonlinear radial glial springs, the nullcline will not be exactly parabolic due to the presence of other nonlinearities in Eq. \ref{general_kr_EL}.

The measure in Eq. \ref{cta} is bounded within $(-1,1)$. For a sinusoidal wave, which is perfectly symmetric, the measure equals zero. For the case of studying this asymmetry in the lobes of the nonlinear BWBM model, we use $\theta$ in lieu of $t$ in Eq. \ref{cta}. Studying this asymmetry for the ADO and the nonlinear BWBM model for a range of tuning parameters, we see that the asymmetry scales as 1.04 $\pm$ 0.05 and the nonlinear BWBM model shows a parameter dependent scaling of 0.61 $\pm$ 0.02 and 0.79 $\pm$ 0.08 for two chosen sets of parameters of 5 lobed and 6 lobed systems respectively (see Fig. \ref{asymmetry panel}f). 

\section{Spatial confinement}\label{tanh_constraint}
A cerebellum does not grow in isolation. It encounters steric effects from the cerebrum, brain-stem, and the skull. The effect of the skull on a growing cerebrum has earlier been studied computationally~\cite{nie2010}. We, therefore, are compelled to explore the effects of steric confinement with the BWBM model, particularly since there may be interplay between the area conservation of the sub-cortex and the imposed steric effects at a radial boundary. More specifically, if the lobes are not allowed to grow radially, they may grow tangentially, making way for sharper folds in the cerebellum. To check for this, we model the steric effects on the developing cerebellum by incorporating a logistic function (1+tanh($x$))/2 into the energy functional, 
\begin{equation}\label{energy_functional_wall}
\begin{split}
E\bigg[r,t,\frac{dt}{d\theta}\bigg]=\int d\theta\bigg\{&k_r(r-r_0)^2-k_t(t-t_0)^2\\
&+\beta\bigg(\frac{dt}{d\theta}\bigg)^2+\frac{K_c}{2}(\text{tanh}[q(r-r_c)])\bigg\}.
\end{split}
\end{equation}
The constants of the logistic function are omitted as they vanish in the resulting Euler-Lagrange equations. Here, $K_c$ is the coupling constant, $q^{-1}$ is the width of the step size of the tanh function and $r_c$ is the radial position of the step. The strength of the steric interaction is taken to be $K_c\;q^2$, a quantity whose dimensions matches those of $k_r$. Renormalizing the parameters and variables by $q^{-1}$ avoids singularities in the Euler-Lagrange equations in the limit of $q\rightarrow0$. Dividing Eq. \ref{energy_functional_wall} by $k_r q^{-2}$, we have,
\begin{equation}
\begin{split}
\tilde{E}\bigg[\tilde{r},\tilde{t},\frac{d\tilde{t}}{d\theta}\bigg] = \int d\theta\bigg\{&(\tilde{r}-\tilde{r}_0)^2-\frac{1}{c}(\tilde{t}-\tilde{t}_0)^2\\
&+\frac{1}{\rho\;c}\bigg(\frac{d\tilde{t}}{d\theta}\bigg)^2+\tilde{K}_c\;\text{tanh}[\tilde{r}-\tilde{r}_c]^2\bigg\},
\end{split}
\end{equation}
where $\tilde{E}:=E/(k_r q^{-2})$, $\tilde{r}:=q r$, $\tilde{t}:=q t$, $\tilde{r}_c:=q r_c$ and $\tilde{K}_c=K_c q^2/k_r$. All parameters and variables are now rendered dimensionless. Given that there is no explicit dependence on $q$, we can be assured that it won't show up in the Euler-Lagrange equation either. The corresponding area conserving constraint is, 
\begin{equation}
\frac{1}{2}\int d\theta(\tilde{r}-\tilde{t})^2=A_0 q^2=\text{dimensionless constant}.
\end{equation}
The Euler-Lagrange equation is, 
\begin{equation}\label{nondim_tanh_circ}
\begin{split}
&\{(1-\epsilon)-\tilde{K}_c\;\text{sech}^2(\tilde{r}-\tilde{r}_c)\;\text{tanh}(\tilde{r}-\tilde{r}_c)\}\left(\frac{d^2 \tilde{r}}{d\theta^2}\right)\\
&+\tilde{K}_c\;\text{sech}^2(\tilde{r}-\tilde{r}_c)\;\{2-3\;\text{sech}^2(\tilde{r}-\tilde{r}_c)\}\left(\frac{d \tilde{r}}{d\theta}\right)^2\\
&+\left\{\rho\epsilon^2 c+(1-\epsilon)\gamma^2\right\}\;\tilde{r}\\
&+\tilde{K}_c\gamma^2\;\text{sech}^2(\tilde{r}-\tilde{r}_c)\\
&+\rho\epsilon\tilde{t_0}- \gamma^2\tilde{r}_0=0.
\end{split}
\end{equation} 
Here $\gamma=\sqrt{\rho(\epsilon c+1)}$. We, again, observe the $(dr/d\theta)^2$ term. However, its coefficient is zero for all $r\neq r_c$, thus effectively localizing its effect. Results of numerical integration in Fig. \ref{tanh panel}a,b shows local flattening of the crests/gyri at contact with the confining wall. Interestingly, there are no nonlocal effects of the confining wall on the lobes, i.e., steric effects do not contribute to the sharpness of the troughs/sulci, at least for the parameters we study. The area conservation on the sub-cortex is not a strong enough constraint to effect the shape of the lobes near the troughs as the sub-cortex can still change its shape. 

\section{A branching hierarchy}\label{hierarchy}

\begin{figure*}[tbh]
\includegraphics[width=1\textwidth]{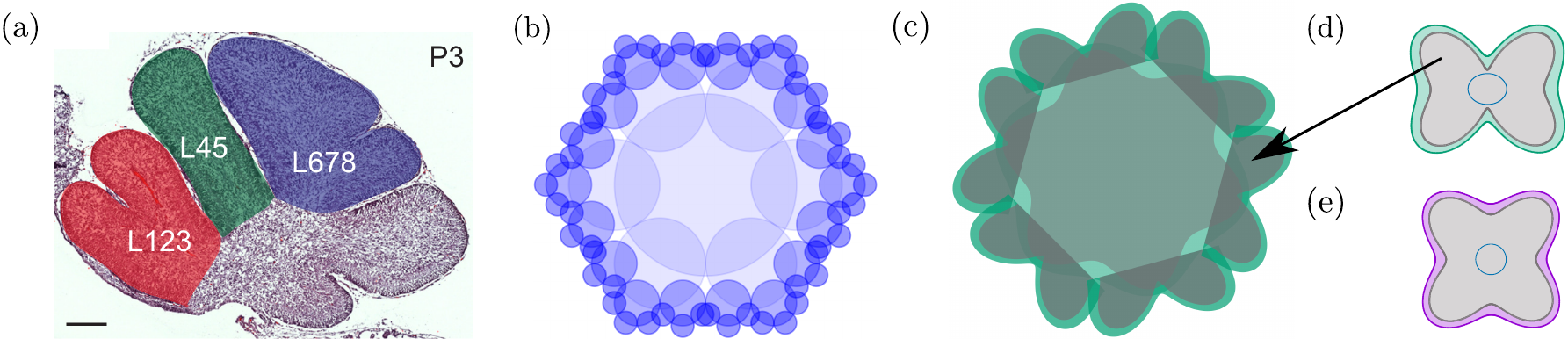}
\caption{\textit{Branching morphogenesis in the cerebellum and length-scale invariance in BWBM foliation.} a) Midline saggital cross-section of mouse cerebellum 3 days post birth. Scale bar is 200 $\mu$m. Reprinted from Fig. 7b of Ref. \cite{lawton2019} published under the terms of the \href{https://creativecommons.org/licenses/by/4.0/}{Creative Commons Attribution}. b) Idealized geometrical example of hierarchy manifests as a fractal.  c) Representation of branching morphogenesis implemented using the geometrical idea of (b) and the 4 lobe BWBM systems with elliptical $r_0$ of (d). Each lobe becomes a sub-system of its own and spawns its own folds. Here, the first and each of the second generation systems are generated numerically with the second generation overlaying the first after a rescaling factor of 0.28. d,e) Folds in (d) develop transversely to the major axis of the elliptical $r_0$ (blue) in contrast to (e) which has a circular $r_0$. \textit{Parameters:} $c=0.0026,\epsilon=0.6,\rho=15.6,\tilde{t}_0=0.58,\tilde{k}_{r_1}=\tilde{k}_{r_2}$ and eccentricity $e=0,06$ respectively for the purple and green 4 lobed systems respectively. The initial condition at $t=0$ is [$\tilde{r},\frac{d\tilde{r}}{d\theta}]=[0.76,0]$. The parameters used for the 6 lobed system is the same as the linear BWBM values in Fig. \ref{expt comparison}. The 4 lobed second generation lobes in (c) are generated using the parameter values corresponding to $e=0.6$ in (b).}\label{hierarchy panel}
\end{figure*}

Mammalian cerebellums are seen to develop folds irrespective of the size of the organ \cite{larsell}. This is in contrast to small mammalian cerebrums which do not develop folds \cite{manyuhina2014,tallinen}. A feature of the linear BWBM model is that it offers an explanation for the length scale independence of the folding morphogenesis in the cerebellum. In the limit of small $\epsilon$ of the linear BWBM model, the number of primary folds $N$ scales as $\sqrt{\rho}$. Thus, in this limit, the number of folds developed in the cerebellum is independent of the size of the cerebellum and is determined solely by the material elastic constants. If the material elastic constants do not dramatically differ across mammalian species, the linear BWBM model can potentially explain the conservation in the number of primary folds. In this section, we discuss how the linear BWBM can be used to describe the branching hierarchy within a given cerebellum.

As the gyri/crests of the cerebellar cortex continue to grow, the sulci/troughs sharpen and become anchored~\cite{sudarov}. The anchoring is due to a combination of the radial glial cells and the pial basement membrane, a thin sheet of extracellular matrix (ECM) made up of collagen, laminin, and other ECM components \cite{pial_membrane_2002}. It is known that basement membranes stick together~\cite{basement_membrane_2014}, so as the troughs sharpen, the pial membrane from each side of the trough begins to come into contact and stick to reinforce the anchoring. The resulting anchoring centers delineate the petal-like projections called lobules \cite{marzban}. 

We hypothesize that when the anchoring centers serve as effective boundaries between the lobules, each lobe becomes its own subsystem. This subsystem then consists of its own subcortex/subcore, all within the encompassing primary cortex/core geometry. Some of these featureless subsystems go on to develop folds of their own to, in turn, generate another generation of subsystems and so on. See Fig. \ref{hierarchy panel}a. In other words, as the cerebellum continues to develop, a branching hierarchy of subsystems emerges. This branching hierarchy is yet another  distinguishing feature of the cerebellum that sets it apart from the cerebrum. It is to be noted that even though cerebellums of all sizes demonstrate folds, smaller cerebellums have fewer hierarchial branchings.

The hierarchical generation of lobules within lobules points to a scale-invariant branching process. Given that within the framework of BWBM, the formation of crests and troughs do not depend on system size, this framework offers a natural description for the hierarchy. As the size of the subsystems decreases with each successive generation, crests and troughs can still form as long as the material properties do not change. The validity of the two dimensional model to describe growth in three dimensions remains valid since the cerebellum retains its cylindrical symmetry during development \cite{lawton2019}.

As an idealized, purely geometric example of the hierarchy, we consider an initial zeroth generation circle. Along its perimeter, six first generation circles are generated. This process can proceed ad infinitum, to generate a fractal structure with a fractal dimension of $\text{log}(3)/\text{log}(2)$.  We show four generations of this hierarchy in Fig. \ref{hierarchy panel}b.  

In the same vein, using the linear BWBM model, we represent the hierarchy of folds in the cerebellum in Fig. \ref{hierarchy panel}c. The `zeroth' generation is the circular $r_0$ (not shown in figure) which generates six first generation foliations. Each lobe formed within consecutive folds is now considered an independent subsystem and sets the length scale for the $r_0$ of the following, second generation of lobes.

Fig. \ref{hierarchy panel}a suggests the free part of the first generation lobe i.e. the part that is not sticking to its neighbors, to be `elliptical' with the major axis of the ellipse being parallel to the outermost exposed edge of the lobe. This is especially evident for the L678 lobe. To accomodate this visual observation in generating second generation folds, we employ a geometric non-linearity in the form of an elliptical $r_0$ with an eccentricity $e$ within the BWBM model. We assume there are no residual stresses and generate a 4 lobed BWBM system as in Fig. \ref{hierarchy panel}d. We then use only the top half of this solution to represent the second generation lobes in Fig. \ref{hierarchy panel}c.

It is also interesting to note that in Fig. \ref{hierarchy panel}d, the most prominent fold occurs transversely to the horizontal major axis of the elliptical $r_0$. This is simply the consequence of the system trying to minimize its energy contribution from the radial glial springs.  For comparison, we show in Fig. \ref{hierarchy panel}e, a four lobed system generated formed from a circular $r_0$.

\section{Discussion}\label{discussion}
Inspired by cerebellar shape development, we study the effects of nonlinear elasticity, steric confinement, and a branching hierarchy within the BWBM model. Exploring the effects of nonlinear elasticity of the fibrous radial glial cells, the interplay between geometry and nonlinearity is seen to give rise to troughs sharper than the troughs obtained in the linear BWBM model and we arrive at an asymmetry between the crests (gyri) and the troughs (sulci). This asymmetry can be understood thus: the relatively slow growth of the sub-cortex is taken into account by demanding area conservation of the sub-cortex in the two-dimensional BWBM model. The associated Lagrange multiplier couples the radius of the cerebellum and the thickness of the cortex. Nonlinear radial springs, in association with this coupling, results in the robust quadratic nonlinear term of the form $(dr/d\theta)^2$ in the shape equation. To illustrate the role of the quadratic nonlinear term in sharpening the sulci, we study the simple harmonic oscillator with the same form of nonlinearity and observe its sufficiency in achieving sharp sulci. Several other nonlinearities emerge in the shape equation including a spatially-varying effective `mass' coefficient. 

The perspective of cerebellum foliation as the action of a nonlinear oscillator can be a useful one given the extensive theoretical studies of such oscillators \cite{nayfeh2008nonlinear,hayashi2014nonlinear}.  For BWBM of the cerebellum, the linear model with constant $r_0$ maps to a forced harmonic oscillator and, for small eccentricities of $r_0$, maps to an unconventional Duffing oscillator. For nonlinear $\tilde{K}_r(r)$, we attempt to understand the corresponding nonlinearity in the context of the assisting-dampening oscillator. We hope the study of cerebellar foliation as a nonlinear oscillator problem continues to be fruitful. In a related work, the existence of a new morphological instability in confined nonlinear elastic sheets was found in the context of a period-doubling bifurcation, exhibiting an analogy with parametric resonance in another nonlinear oscillator~\cite{brau2011}.

The period-doubling hierarchy found in the Ref. ~\cite{brau2011} is very different from the new hierarchy found here. The hierarchy found here is one due to boundary conditions in the form of anchoring centers to create sub-regions, or sub-systems, from which the same type of scale-invariant foliation emerges, at least in the limit of small $\epsilon$.  Had the foliation mechanism not been scale-invariant in any limit, such as with a purely elastic system, the smaller sub-lobes would soon become featureless as the number of foliations depend linearly on the perimeter of the sub-system.  Within BWBM, therefore, we have identified a new scale-invariant branching morphogenesis mechanism. It is not yet clear how generic this new branching hierarchy mechanism is in terms of moving beyond the cerebellum. Ref.~\cite{engstrom2018} addressed potential applications of BWBM to two-dimensional brain organoids~\cite{karzbrun2018} and the developing retina \cite{kol18}. 

Referring again to Fig. \ref{hierarchy panel}(a), one of the second generation sublobes labelled L45 does not branch. Perhaps the material properties are altered in this sublobe so that features do not form. For sublobe L678, the two new sub-sublobes are not similar in size.  This could be due to changes in curvature of the confinement from growing, surrounding tissue.  So far, we have only addressed steric, static confinement.  Certainly, such variabilities from sublobe to sublobe can be explored in less idealized conditions.

We note that within the context of purely elasticity theory, an explanation for the thicker-sulci/thinner-gyri of the developing cerebellar cortex was recently achieved by the addition of surface tension \cite{riccobelli}.  While more energetic contributions can certainly be added to either the purely elastic model or to the BWBM model, the recent experimental observations needs to be incorporated in the modeling -- the cells in the cortex are motile with cellular rearrangements on the time scale of minutes ~\cite{lawton2019} and the cerebellum is under tension as it develops (as opposed to compression). These observations render a purely elastic model suspect. However, the differential growth between the cortex and sub-cortex remains central in both classes of models. To make progress, we need further experimental falsification tests to rule out classes of models.

Finally, given our focus on the cerebellum here, one is led to wonder whether some form of BWBM is applicable to the cerebrum.  As mentioned previously, the cerebrum and the cerebellum have rather different morphologies.  In terms of development in the cerebellum, the predominant growth of the cells is in the cortex~\cite{sudarov}. Many such cells migrate inward to become part of the core of the cerebellum.  In the cerebrum, much of the cell proliferation is in the core and the progenitor cells migrate outward to become part of the cortex \cite{rakic2009,molnar2019} . In this sense, the two organs are inverse to each other.  Given the presence of motile cells in the developing cerebrum, one may wonder whether a purely elastic approach to the developing brain is reasonable. Without a doubt, the time scales of cell migration decide the contest between elasticity and fluidity. Under the framework of liquid crystals, earlier work on the developing cerebrum arrived at a prediction for the bending modulus of the cortex~\cite{manyuhina2014}. This approach was based on a revised view of the axonal tension model for the developing cerebrum \cite{van1997,van2020}. These novel approaches have the potential to build new inroads in quantitative biology. 

MCG acknowledges useful discussions with Manu Mannattil. JMS acknowledges financial support from NSF-DMR-1832002 and an Isaac Newton Award from the DoD. 

\bibliography{References}

\begin{thebibliography}{49}%
\makeatletter
\providecommand \@ifxundefined [1]{%
 \@ifx{#1\undefined}
}%
\providecommand \@ifnum [1]{%
 \ifnum #1\expandafter \@firstoftwo
 \else \expandafter \@secondoftwo
 \fi
}%
\providecommand \@ifx [1]{%
 \ifx #1\expandafter \@firstoftwo
 \else \expandafter \@secondoftwo
 \fi
}%
\providecommand \natexlab [1]{#1}%
\providecommand \enquote  [1]{``#1''}%
\providecommand \bibnamefont  [1]{#1}%
\providecommand \bibfnamefont [1]{#1}%
\providecommand \citenamefont [1]{#1}%
\providecommand \href@noop [0]{\@secondoftwo}%
\providecommand \href [0]{\begingroup \@sanitize@url \@href}%
\providecommand \@href[1]{\@@startlink{#1}\@@href}%
\providecommand \@@href[1]{\endgroup#1\@@endlink}%
\providecommand \@sanitize@url [0]{\catcode `\\12\catcode `\$12\catcode
  `\&12\catcode `\#12\catcode `\^12\catcode `\_12\catcode `\%12\relax}%
\providecommand \@@startlink[1]{}%
\providecommand \@@endlink[0]{}%
\providecommand \url  [0]{\begingroup\@sanitize@url \@url }%
\providecommand \@url [1]{\endgroup\@href {#1}{\urlprefix }}%
\providecommand \urlprefix  [0]{URL }%
\providecommand \Eprint [0]{\href }%
\providecommand \doibase [0]{http://dx.doi.org/}%
\providecommand \selectlanguage [0]{\@gobble}%
\providecommand \bibinfo  [0]{\@secondoftwo}%
\providecommand \bibfield  [0]{\@secondoftwo}%
\providecommand \translation [1]{[#1]}%
\providecommand \BibitemOpen [0]{}%
\providecommand \bibitemStop [0]{}%
\providecommand \bibitemNoStop [0]{.\EOS\space}%
\providecommand \EOS [0]{\spacefactor3000\relax}%
\providecommand \BibitemShut  [1]{\csname bibitem#1\endcsname}%
\let\auto@bib@innerbib\@empty
\bibitem [{\citenamefont {Ben~Amar}\ \emph {et~al.}(2019)\citenamefont
  {Ben~Amar}, \citenamefont {Nassoy},\ and\ \citenamefont
  {LeGoff}}]{ben_amar_2019}%
  \BibitemOpen
  \bibfield  {author} {\bibinfo {author} {\bibfnamefont {M.}~\bibnamefont
  {Ben~Amar}}, \bibinfo {author} {\bibfnamefont {P.}~\bibnamefont {Nassoy}}, \
  and\ \bibinfo {author} {\bibfnamefont {L.}~\bibnamefont {LeGoff}},\
  }\href@noop {} {\bibfield  {journal} {\bibinfo  {journal} {Philosophical
  Transactions of the Royal Society A}\ }\textbf {\bibinfo {volume} {377}},\
  \bibinfo {pages} {20180070} (\bibinfo {year} {2019})}\BibitemShut {NoStop}%
\bibitem [{\citenamefont {Nelson}(2016)}]{nelson2016buckling}%
  \BibitemOpen
  \bibfield  {author} {\bibinfo {author} {\bibfnamefont {C.~M.}\ \bibnamefont
  {Nelson}},\ }\href@noop {} {\bibfield  {journal} {\bibinfo  {journal}
  {Journal of Biomechanical Engineering}\ }\textbf {\bibinfo {volume} {138}}
  (\bibinfo {year} {2016})}\BibitemShut {NoStop}%
\bibitem [{\citenamefont {Budday}\ \emph {et~al.}(2015)\citenamefont {Budday},
  \citenamefont {Steinmann~III},\ and\ \citenamefont
  {Kuhl}}]{budday2015physical}%
  \BibitemOpen
  \bibfield  {author} {\bibinfo {author} {\bibfnamefont {S.}~\bibnamefont
  {Budday}}, \bibinfo {author} {\bibfnamefont {P.}~\bibnamefont
  {Steinmann~III}}, \ and\ \bibinfo {author} {\bibfnamefont {E.}~\bibnamefont
  {Kuhl}},\ }\href@noop {} {\bibfield  {journal} {\bibinfo  {journal}
  {Frontiers in Cellular Neuroscience}\ }\textbf {\bibinfo {volume} {9}},\
  \bibinfo {pages} {257} (\bibinfo {year} {2015})}\BibitemShut {NoStop}%
\bibitem [{\citenamefont {Richman}\ \emph {et~al.}(1975)\citenamefont
  {Richman}, \citenamefont {Stewart}, \citenamefont {Hutchinson},\ and\
  \citenamefont {Caviness}}]{richman1975mechanical}%
  \BibitemOpen
  \bibfield  {author} {\bibinfo {author} {\bibfnamefont {D.~P.}\ \bibnamefont
  {Richman}}, \bibinfo {author} {\bibfnamefont {R.~M.}\ \bibnamefont
  {Stewart}}, \bibinfo {author} {\bibfnamefont {J.~W.}\ \bibnamefont
  {Hutchinson}}, \ and\ \bibinfo {author} {\bibfnamefont {V.~S.}\ \bibnamefont
  {Caviness}},\ }\href@noop {} {\bibfield  {journal} {\bibinfo  {journal}
  {Science}\ }\textbf {\bibinfo {volume} {189}},\ \bibinfo {pages} {18}
  (\bibinfo {year} {1975})}\BibitemShut {NoStop}%
\bibitem [{\citenamefont {Raghavan}\ \emph {et~al.}(1997)\citenamefont
  {Raghavan}, \citenamefont {Lawton}, \citenamefont {Ranjan},\ and\
  \citenamefont {Viswanathan}}]{raghavan1997continuum}%
  \BibitemOpen
  \bibfield  {author} {\bibinfo {author} {\bibfnamefont {R.}~\bibnamefont
  {Raghavan}}, \bibinfo {author} {\bibfnamefont {W.}~\bibnamefont {Lawton}},
  \bibinfo {author} {\bibfnamefont {S.}~\bibnamefont {Ranjan}}, \ and\ \bibinfo
  {author} {\bibfnamefont {R.}~\bibnamefont {Viswanathan}},\ }\href@noop {}
  {\bibfield  {journal} {\bibinfo  {journal} {Journal of Theoretical Biology}\
  }\textbf {\bibinfo {volume} {187}},\ \bibinfo {pages} {285} (\bibinfo {year}
  {1997})}\BibitemShut {NoStop}%
\bibitem [{\citenamefont {Bayly}\ \emph {et~al.}(2013)\citenamefont {Bayly},
  \citenamefont {Okamoto}, \citenamefont {Xu}, \citenamefont {Shi},\ and\
  \citenamefont {Taber}}]{bayly2013cortical}%
  \BibitemOpen
  \bibfield  {author} {\bibinfo {author} {\bibfnamefont {P.}~\bibnamefont
  {Bayly}}, \bibinfo {author} {\bibfnamefont {R.}~\bibnamefont {Okamoto}},
  \bibinfo {author} {\bibfnamefont {G.}~\bibnamefont {Xu}}, \bibinfo {author}
  {\bibfnamefont {Y.}~\bibnamefont {Shi}}, \ and\ \bibinfo {author}
  {\bibfnamefont {L.}~\bibnamefont {Taber}},\ }\href@noop {} {\bibfield
  {journal} {\bibinfo  {journal} {Physical Biology}\ }\textbf {\bibinfo
  {volume} {10}},\ \bibinfo {pages} {016005} (\bibinfo {year}
  {2013})}\BibitemShut {NoStop}%
\bibitem [{\citenamefont {Tallinen}\ \emph {et~al.}(2014)\citenamefont
  {Tallinen}, \citenamefont {Chung}, \citenamefont {Biggins},\ and\
  \citenamefont {Mahadevan}}]{tallinen}%
  \BibitemOpen
  \bibfield  {author} {\bibinfo {author} {\bibfnamefont {T.}~\bibnamefont
  {Tallinen}}, \bibinfo {author} {\bibfnamefont {J.~Y.}\ \bibnamefont {Chung}},
  \bibinfo {author} {\bibfnamefont {J.~S.}\ \bibnamefont {Biggins}}, \ and\
  \bibinfo {author} {\bibfnamefont {L.}~\bibnamefont {Mahadevan}},\ }\href@noop
  {} {\bibfield  {journal} {\bibinfo  {journal} {Proceedings of the National
  Academy of Sciences}\ }\textbf {\bibinfo {volume} {111}},\ \bibinfo {pages}
  {12667} (\bibinfo {year} {2014})}\BibitemShut {NoStop}%
\bibitem [{\citenamefont {Tallinen}\ \emph {et~al.}(2016)\citenamefont
  {Tallinen}, \citenamefont {Chung}, \citenamefont {Rousseau}, \citenamefont
  {Girard}, \citenamefont {Lef{\`e}vre},\ and\ \citenamefont
  {Mahadevan}}]{tallinen2016growth}%
  \BibitemOpen
  \bibfield  {author} {\bibinfo {author} {\bibfnamefont {T.}~\bibnamefont
  {Tallinen}}, \bibinfo {author} {\bibfnamefont {J.~Y.}\ \bibnamefont {Chung}},
  \bibinfo {author} {\bibfnamefont {F.}~\bibnamefont {Rousseau}}, \bibinfo
  {author} {\bibfnamefont {N.}~\bibnamefont {Girard}}, \bibinfo {author}
  {\bibfnamefont {J.}~\bibnamefont {Lef{\`e}vre}}, \ and\ \bibinfo {author}
  {\bibfnamefont {L.}~\bibnamefont {Mahadevan}},\ }\href@noop {} {\bibfield
  {journal} {\bibinfo  {journal} {Nature Physics}\ }\textbf {\bibinfo {volume}
  {12}},\ \bibinfo {pages} {588} (\bibinfo {year} {2016})}\BibitemShut
  {NoStop}%
\bibitem [{\citenamefont {Hannezo}\ \emph {et~al.}(2011)\citenamefont
  {Hannezo}, \citenamefont {Prost},\ and\ \citenamefont
  {Joanny}}]{hannezo2011instabilities}%
  \BibitemOpen
  \bibfield  {author} {\bibinfo {author} {\bibfnamefont {E.}~\bibnamefont
  {Hannezo}}, \bibinfo {author} {\bibfnamefont {J.}~\bibnamefont {Prost}}, \
  and\ \bibinfo {author} {\bibfnamefont {J.-F.}\ \bibnamefont {Joanny}},\
  }\href@noop {} {\bibfield  {journal} {\bibinfo  {journal} {Physical Review
  Letters}\ }\textbf {\bibinfo {volume} {107}},\ \bibinfo {pages} {078104}
  (\bibinfo {year} {2011})}\BibitemShut {NoStop}%
\bibitem [{\citenamefont {Shyer}\ \emph {et~al.}(2013)\citenamefont {Shyer},
  \citenamefont {Tallinen}, \citenamefont {Nerurkar}, \citenamefont {Wei},
  \citenamefont {Gil}, \citenamefont {Kaplan}, \citenamefont {Tabin},\ and\
  \citenamefont {Mahadevan}}]{shyer2013villification}%
  \BibitemOpen
  \bibfield  {author} {\bibinfo {author} {\bibfnamefont {A.~E.}\ \bibnamefont
  {Shyer}}, \bibinfo {author} {\bibfnamefont {T.}~\bibnamefont {Tallinen}},
  \bibinfo {author} {\bibfnamefont {N.~L.}\ \bibnamefont {Nerurkar}}, \bibinfo
  {author} {\bibfnamefont {Z.}~\bibnamefont {Wei}}, \bibinfo {author}
  {\bibfnamefont {E.~S.}\ \bibnamefont {Gil}}, \bibinfo {author} {\bibfnamefont
  {D.~L.}\ \bibnamefont {Kaplan}}, \bibinfo {author} {\bibfnamefont {C.~J.}\
  \bibnamefont {Tabin}}, \ and\ \bibinfo {author} {\bibfnamefont
  {L.}~\bibnamefont {Mahadevan}},\ }\href@noop {} {\bibfield  {journal}
  {\bibinfo  {journal} {Science}\ }\textbf {\bibinfo {volume} {342}},\ \bibinfo
  {pages} {212} (\bibinfo {year} {2013})}\BibitemShut {NoStop}%
\bibitem [{\citenamefont {Wiggs}\ \emph {et~al.}(1997)\citenamefont {Wiggs},
  \citenamefont {Hrousis}, \citenamefont {Drazen},\ and\ \citenamefont
  {Kamm}}]{wiggs1997mechanism}%
  \BibitemOpen
  \bibfield  {author} {\bibinfo {author} {\bibfnamefont {B.~R.}\ \bibnamefont
  {Wiggs}}, \bibinfo {author} {\bibfnamefont {C.~A.}\ \bibnamefont {Hrousis}},
  \bibinfo {author} {\bibfnamefont {J.~M.}\ \bibnamefont {Drazen}}, \ and\
  \bibinfo {author} {\bibfnamefont {R.~D.}\ \bibnamefont {Kamm}},\ }\href@noop
  {} {\bibfield  {journal} {\bibinfo  {journal} {Journal of Applied
  Physiology}\ }\textbf {\bibinfo {volume} {83}},\ \bibinfo {pages} {1814}
  (\bibinfo {year} {1997})}\BibitemShut {NoStop}%
\bibitem [{\citenamefont {Li}\ \emph {et~al.}(2011)\citenamefont {Li},
  \citenamefont {Cao}, \citenamefont {Feng},\ and\ \citenamefont
  {Gao}}]{li2011surface}%
  \BibitemOpen
  \bibfield  {author} {\bibinfo {author} {\bibfnamefont {B.}~\bibnamefont
  {Li}}, \bibinfo {author} {\bibfnamefont {Y.-P.}\ \bibnamefont {Cao}},
  \bibinfo {author} {\bibfnamefont {X.-Q.}\ \bibnamefont {Feng}}, \ and\
  \bibinfo {author} {\bibfnamefont {H.}~\bibnamefont {Gao}},\ }\href@noop {}
  {\bibfield  {journal} {\bibinfo  {journal} {Journal of the Mechanics and
  Physics of Solids}\ }\textbf {\bibinfo {volume} {59}},\ \bibinfo {pages}
  {758} (\bibinfo {year} {2011})}\BibitemShut {NoStop}%
\bibitem [{\citenamefont {Osborn}(2008)}]{osborn2008model}%
  \BibitemOpen
  \bibfield  {author} {\bibinfo {author} {\bibfnamefont {J.~W.}\ \bibnamefont
  {Osborn}},\ }\href@noop {} {\bibfield  {journal} {\bibinfo  {journal}
  {Journal of Theoretical Biology}\ }\textbf {\bibinfo {volume} {255}},\
  \bibinfo {pages} {338} (\bibinfo {year} {2008})}\BibitemShut {NoStop}%
\bibitem [{\citenamefont {Shyer}\ \emph {et~al.}(2017)\citenamefont {Shyer},
  \citenamefont {Rodrigues}, \citenamefont {Schroeder}, \citenamefont
  {Kassianidou}, \citenamefont {Kumar},\ and\ \citenamefont
  {Harland}}]{shyer2017emergent}%
  \BibitemOpen
  \bibfield  {author} {\bibinfo {author} {\bibfnamefont {A.~E.}\ \bibnamefont
  {Shyer}}, \bibinfo {author} {\bibfnamefont {A.~R.}\ \bibnamefont
  {Rodrigues}}, \bibinfo {author} {\bibfnamefont {G.~G.}\ \bibnamefont
  {Schroeder}}, \bibinfo {author} {\bibfnamefont {E.}~\bibnamefont
  {Kassianidou}}, \bibinfo {author} {\bibfnamefont {S.}~\bibnamefont {Kumar}},
  \ and\ \bibinfo {author} {\bibfnamefont {R.~M.}\ \bibnamefont {Harland}},\
  }\href@noop {} {\bibfield  {journal} {\bibinfo  {journal} {Science}\ }\textbf
  {\bibinfo {volume} {357}},\ \bibinfo {pages} {811} (\bibinfo {year}
  {2017})}\BibitemShut {NoStop}%
\bibitem [{\citenamefont {Lejeune}\ \emph {et~al.}(2016)\citenamefont
  {Lejeune}, \citenamefont {Javili}, \citenamefont {Weickenmeier},
  \citenamefont {Kuhl},\ and\ \citenamefont {Linder}}]{lejeune}%
  \BibitemOpen
  \bibfield  {author} {\bibinfo {author} {\bibfnamefont {E.}~\bibnamefont
  {Lejeune}}, \bibinfo {author} {\bibfnamefont {A.}~\bibnamefont {Javili}},
  \bibinfo {author} {\bibfnamefont {J.}~\bibnamefont {Weickenmeier}}, \bibinfo
  {author} {\bibfnamefont {E.}~\bibnamefont {Kuhl}}, \ and\ \bibinfo {author}
  {\bibfnamefont {C.}~\bibnamefont {Linder}},\ }\href@noop {} {\bibfield
  {journal} {\bibinfo  {journal} {Soft Matter}\ }\textbf {\bibinfo {volume}
  {12}},\ \bibinfo {pages} {5613} (\bibinfo {year} {2016})}\BibitemShut
  {NoStop}%
\bibitem [{\citenamefont {Lejeune}\ \emph {et~al.}(2019)\citenamefont
  {Lejeune}, \citenamefont {Dortdivanlioglu}, \citenamefont {Kuhl},\ and\
  \citenamefont {Linder}}]{lejeune2019}%
  \BibitemOpen
  \bibfield  {author} {\bibinfo {author} {\bibfnamefont {E.}~\bibnamefont
  {Lejeune}}, \bibinfo {author} {\bibfnamefont {B.}~\bibnamefont
  {Dortdivanlioglu}}, \bibinfo {author} {\bibfnamefont {E.}~\bibnamefont
  {Kuhl}}, \ and\ \bibinfo {author} {\bibfnamefont {C.}~\bibnamefont
  {Linder}},\ }\href@noop {} {\bibfield  {journal} {\bibinfo  {journal} {Soft
  Matter}\ }\textbf {\bibinfo {volume} {15}},\ \bibinfo {pages} {2204}
  (\bibinfo {year} {2019})}\BibitemShut {NoStop}%
\bibitem [{\citenamefont {Mongera}\ and\ \citenamefont
  {et~al.}(2018)}]{mongera_2018}%
  \BibitemOpen
  \bibfield  {author} {\bibinfo {author} {\bibfnamefont {A.}~\bibnamefont
  {Mongera}}\ and\ \bibinfo {author} {\bibnamefont {et~al.}},\ }\href@noop {}
  {\bibfield  {journal} {\bibinfo  {journal} {Nature}\ }\textbf {\bibinfo
  {volume} {561}},\ \bibinfo {pages} {401} (\bibinfo {year}
  {2018})}\BibitemShut {NoStop}%
\bibitem [{\citenamefont {Jain}\ and\ \citenamefont {et~al}(2020)}]{jain_2020}%
  \BibitemOpen
  \bibfield  {author} {\bibinfo {author} {\bibfnamefont {A.}~\bibnamefont
  {Jain}}\ and\ \bibinfo {author} {\bibnamefont {et~al}},\ }\href@noop {}
  {\bibfield  {journal} {\bibinfo  {journal} {Nature Communications}\ }\textbf
  {\bibinfo {volume} {11}},\ \bibinfo {pages} {5604} (\bibinfo {year}
  {2020})}\BibitemShut {NoStop}%
\bibitem [{\citenamefont {Engstrom}\ \emph {et~al.}(2018)\citenamefont
  {Engstrom}, \citenamefont {Zhang}, \citenamefont {Lawton}, \citenamefont
  {Joyner},\ and\ \citenamefont {Schwarz}}]{engstrom2018}%
  \BibitemOpen
  \bibfield  {author} {\bibinfo {author} {\bibfnamefont {T.}~\bibnamefont
  {Engstrom}}, \bibinfo {author} {\bibfnamefont {T.}~\bibnamefont {Zhang}},
  \bibinfo {author} {\bibfnamefont {A.}~\bibnamefont {Lawton}}, \bibinfo
  {author} {\bibfnamefont {A.}~\bibnamefont {Joyner}}, \ and\ \bibinfo {author}
  {\bibfnamefont {J.~M.}\ \bibnamefont {Schwarz}},\ }\href@noop {} {\bibfield
  {journal} {\bibinfo  {journal} {Physical Review X}\ }\textbf {\bibinfo
  {volume} {8}},\ \bibinfo {pages} {041053} (\bibinfo {year}
  {2018})}\BibitemShut {NoStop}%
\bibitem [{\citenamefont {Larsell}(1967)}]{larsell}%
  \BibitemOpen
  \bibfield  {author} {\bibinfo {author} {\bibfnamefont {O.}~\bibnamefont
  {Larsell}},\ }\href@noop {} {\emph {\bibinfo {title} {The Comparative Anatomy
  and Histology of the Cerebellum}}}\ (\bibinfo  {publisher} {University of
  Minnesota Press},\ \bibinfo {year} {1967})\BibitemShut {NoStop}%
\bibitem [{\citenamefont {Herculano-Houzel}(2009)}]{herculano-houzel_2009}%
  \BibitemOpen
  \bibfield  {author} {\bibinfo {author} {\bibfnamefont {S.}~\bibnamefont
  {Herculano-Houzel}},\ }\href@noop {} {\bibfield  {journal} {\bibinfo
  {journal} {Front. Hum. Neurosci.}\ }\textbf {\bibinfo {volume} {3}},\
  \bibinfo {pages} {31} (\bibinfo {year} {2009})}\BibitemShut {NoStop}%
\bibitem [{\citenamefont {Lawton}\ \emph {et~al.}(2019)\citenamefont {Lawton},
  \citenamefont {Engstrom}, \citenamefont {Rohrbach}, \citenamefont {Omura},
  \citenamefont {Turnbull}, \citenamefont {Mamou}, \citenamefont {Zhang},
  \citenamefont {Schwarz},\ and\ \citenamefont {Joyner}}]{lawton2019}%
  \BibitemOpen
  \bibfield  {author} {\bibinfo {author} {\bibfnamefont {A.~K.}\ \bibnamefont
  {Lawton}}, \bibinfo {author} {\bibfnamefont {T.}~\bibnamefont {Engstrom}},
  \bibinfo {author} {\bibfnamefont {D.}~\bibnamefont {Rohrbach}}, \bibinfo
  {author} {\bibfnamefont {M.}~\bibnamefont {Omura}}, \bibinfo {author}
  {\bibfnamefont {D.~H.}\ \bibnamefont {Turnbull}}, \bibinfo {author}
  {\bibfnamefont {J.}~\bibnamefont {Mamou}}, \bibinfo {author} {\bibfnamefont
  {T.}~\bibnamefont {Zhang}}, \bibinfo {author} {\bibfnamefont {J.~M.}\
  \bibnamefont {Schwarz}}, \ and\ \bibinfo {author} {\bibfnamefont {A.~L.}\
  \bibnamefont {Joyner}},\ }\href@noop {} {\bibfield  {journal} {\bibinfo
  {journal} {Elife}\ }\textbf {\bibinfo {volume} {8}},\ \bibinfo {pages}
  {e45019} (\bibinfo {year} {2019})}\BibitemShut {NoStop}%
\bibitem [{\citenamefont {Fernandez}\ \emph {et~al.}(2006)\citenamefont
  {Fernandez}, \citenamefont {Pullarkar},\ and\ \citenamefont
  {Ott}}]{fernandez_2006}%
  \BibitemOpen
  \bibfield  {author} {\bibinfo {author} {\bibfnamefont {P.}~\bibnamefont
  {Fernandez}}, \bibinfo {author} {\bibfnamefont {P.~A.}\ \bibnamefont
  {Pullarkar}}, \ and\ \bibinfo {author} {\bibfnamefont {A.}~\bibnamefont
  {Ott}},\ }\href@noop {} {\bibfield  {journal} {\bibinfo  {journal} {Biophys.
  J.}\ }\textbf {\bibinfo {volume} {90}},\ \bibinfo {pages} {3796} (\bibinfo
  {year} {2006})}\BibitemShut {NoStop}%
\bibitem [{\citenamefont {Varner}\ and\ \citenamefont
  {Nelson}(2014)}]{varner_2014}%
  \BibitemOpen
  \bibfield  {author} {\bibinfo {author} {\bibfnamefont {V.~D.}\ \bibnamefont
  {Varner}}\ and\ \bibinfo {author} {\bibfnamefont {C.~M.}\ \bibnamefont
  {Nelson}},\ }\href@noop {} {\bibfield  {journal} {\bibinfo  {journal}
  {Development}\ }\textbf {\bibinfo {volume} {141}},\ \bibinfo {pages} {2750}
  (\bibinfo {year} {2014})}\BibitemShut {NoStop}%
\bibitem [{\citenamefont {Hannezo}\ \emph {et~al.}(2017)\citenamefont
  {Hannezo}, \citenamefont {Scheele}, \citenamefont {Moad}, \citenamefont
  {Drogo}, \citenamefont {Heer}, \citenamefont {Sampongna}, \citenamefont {van
  Rheenen},\ and\ \citenamefont {Simons}}]{hannezo_2017}%
  \BibitemOpen
  \bibfield  {author} {\bibinfo {author} {\bibfnamefont {E.}~\bibnamefont
  {Hannezo}}, \bibinfo {author} {\bibfnamefont {C.~L. G.~J.}\ \bibnamefont
  {Scheele}}, \bibinfo {author} {\bibfnamefont {M.}~\bibnamefont {Moad}},
  \bibinfo {author} {\bibfnamefont {N.}~\bibnamefont {Drogo}}, \bibinfo
  {author} {\bibfnamefont {R.}~\bibnamefont {Heer}}, \bibinfo {author}
  {\bibfnamefont {R.~V.}\ \bibnamefont {Sampongna}}, \bibinfo {author}
  {\bibfnamefont {J.}~\bibnamefont {van Rheenen}}, \ and\ \bibinfo {author}
  {\bibfnamefont {B.~D.}\ \bibnamefont {Simons}},\ }\href@noop {} {\bibfield
  {journal} {\bibinfo  {journal} {Cell}\ }\textbf {\bibinfo {volume} {171}},\
  \bibinfo {pages} {242} (\bibinfo {year} {2017})}\BibitemShut {NoStop}%
\bibitem [{\citenamefont {Lagrange}\ \emph {et~al.}(2016)\citenamefont
  {Lagrange}, \citenamefont {Jim{\'e}nez}, \citenamefont {Terwagne},
  \citenamefont {Brojan},\ and\ \citenamefont {Reis}}]{lagrange2016wrinkling}%
  \BibitemOpen
  \bibfield  {author} {\bibinfo {author} {\bibfnamefont {R.}~\bibnamefont
  {Lagrange}}, \bibinfo {author} {\bibfnamefont {F.~L.}\ \bibnamefont
  {Jim{\'e}nez}}, \bibinfo {author} {\bibfnamefont {D.}~\bibnamefont
  {Terwagne}}, \bibinfo {author} {\bibfnamefont {M.}~\bibnamefont {Brojan}}, \
  and\ \bibinfo {author} {\bibfnamefont {P.}~\bibnamefont {Reis}},\ }\href@noop
  {} {\bibfield  {journal} {\bibinfo  {journal} {Journal of the Mechanics and
  Physics of Solids}\ }\textbf {\bibinfo {volume} {89}},\ \bibinfo {pages} {77}
  (\bibinfo {year} {2016})}\BibitemShut {NoStop}%
\bibitem [{\citenamefont {Marzban}(2017)}]{marzban}%
  \BibitemOpen
  \bibfield  {author} {\bibinfo {author} {\bibfnamefont {H.}~\bibnamefont
  {Marzban}},\ }\href@noop {} {\emph {\bibinfo {title} {Development of the
  Cerebellum from Molecular Aspects to Diseases}}}\ (\bibinfo  {publisher}
  {Springer},\ \bibinfo {year} {2017})\BibitemShut {NoStop}%
\bibitem [{\citenamefont {Storm}\ \emph {et~al.}(2005)\citenamefont {Storm},
  \citenamefont {Pastore}, \citenamefont {MacKintosh}, \citenamefont
  {Lubensky},\ and\ \citenamefont {A}}]{storm_2005}%
  \BibitemOpen
  \bibfield  {author} {\bibinfo {author} {\bibfnamefont {C.}~\bibnamefont
  {Storm}}, \bibinfo {author} {\bibfnamefont {J.~J.}\ \bibnamefont {Pastore}},
  \bibinfo {author} {\bibfnamefont {F.~C.}\ \bibnamefont {MacKintosh}},
  \bibinfo {author} {\bibfnamefont {T.~C.}\ \bibnamefont {Lubensky}}, \ and\
  \bibinfo {author} {\bibfnamefont {J.~P.}\ \bibnamefont {A}},\ }\href@noop {}
  {\bibfield  {journal} {\bibinfo  {journal} {Nature}\ }\textbf {\bibinfo
  {volume} {435}},\ \bibinfo {pages} {195} (\bibinfo {year}
  {2005})}\BibitemShut {NoStop}%
\bibitem [{\citenamefont {Davis}(1961)}]{davis1961}%
  \BibitemOpen
  \bibfield  {author} {\bibinfo {author} {\bibfnamefont {H.~T.}\ \bibnamefont
  {Davis}},\ }\href@noop {} {\emph {\bibinfo {title} {Introduction to Nonlinear
  Differential and Integral Equations}}}\ (\bibinfo  {publisher} {US Government
  Printing Office},\ \bibinfo {year} {1961})\BibitemShut {NoStop}%
\bibitem [{\citenamefont {Ermentrout}(2002)}]{ermentrout}%
  \BibitemOpen
  \bibfield  {author} {\bibinfo {author} {\bibfnamefont {B.}~\bibnamefont
  {Ermentrout}},\ }\href@noop {} {\emph {\bibinfo {title} {Simulating,
  Analyzing, and Animating Dynamical Systems: A Guide to XPPAUT for Researchers
  and Students}}}\ (\bibinfo  {publisher} {SIAM},\ \bibinfo {year}
  {2002})\BibitemShut {NoStop}%
\bibitem [{\citenamefont {Goldstein}\ \emph {et~al.}(2002)\citenamefont
  {Goldstein}, \citenamefont {Poole},\ and\ \citenamefont {Safko}}]{goldstein}%
  \BibitemOpen
  \bibfield  {author} {\bibinfo {author} {\bibfnamefont {H.}~\bibnamefont
  {Goldstein}}, \bibinfo {author} {\bibfnamefont {C.}~\bibnamefont {Poole}}, \
  and\ \bibinfo {author} {\bibfnamefont {J.}~\bibnamefont {Safko}},\
  }\href@noop {} {\emph {\bibinfo {title} {Classical mechanics}}}\ (\bibinfo
  {publisher} {Addison Wesley},\ \bibinfo {year} {2002})\BibitemShut {NoStop}%
\bibitem [{\citenamefont {Strogatz}(2018)}]{strogatz}%
  \BibitemOpen
  \bibfield  {author} {\bibinfo {author} {\bibfnamefont {S.~H.}\ \bibnamefont
  {Strogatz}},\ }\href@noop {} {\emph {\bibinfo {title} {Nonlinear Dynamics and
  Chaos: With Applications to Physics, Biology, Chemistry, and Engineering}}}\
  (\bibinfo  {publisher} {CRC Press},\ \bibinfo {year} {2018})\BibitemShut
  {NoStop}%
\bibitem [{\citenamefont {Jordan}\ \emph {et~al.}(2007)\citenamefont {Jordan},
  \citenamefont {Smith}, \citenamefont {Smith} \emph {et~al.}}]{jordan2007}%
  \BibitemOpen
  \bibfield  {author} {\bibinfo {author} {\bibfnamefont {D.}~\bibnamefont
  {Jordan}}, \bibinfo {author} {\bibfnamefont {P.}~\bibnamefont {Smith}},
  \bibinfo {author} {\bibfnamefont {P.}~\bibnamefont {Smith}},  \emph
  {et~al.},\ }\href@noop {} {\emph {\bibinfo {title} {Nonlinear Ordinary
  Differential Equations: An Introduction for Scientists and engineers}}},\
  Vol.~\bibinfo {volume} {10}\ (\bibinfo  {publisher} {Oxford University Press
  on Demand},\ \bibinfo {year} {2007})\BibitemShut {NoStop}%
\bibitem [{\citenamefont {Stokes}(1880)}]{stokes}%
  \BibitemOpen
  \bibfield  {author} {\bibinfo {author} {\bibfnamefont {G.~G.}\ \bibnamefont
  {Stokes}},\ }\href@noop {} {\bibfield  {journal} {\bibinfo  {journal}
  {Transactions of the Cambridge Philosophical Society}\ } (\bibinfo {year}
  {1880})}\BibitemShut {NoStop}%
\bibitem [{\citenamefont {Nie}\ \emph {et~al.}(2010)\citenamefont {Nie},
  \citenamefont {Guo}, \citenamefont {Li}, \citenamefont {Faraco},
  \citenamefont {Miller},\ and\ \citenamefont {Liu}}]{nie2010}%
  \BibitemOpen
  \bibfield  {author} {\bibinfo {author} {\bibfnamefont {J.}~\bibnamefont
  {Nie}}, \bibinfo {author} {\bibfnamefont {L.}~\bibnamefont {Guo}}, \bibinfo
  {author} {\bibfnamefont {G.}~\bibnamefont {Li}}, \bibinfo {author}
  {\bibfnamefont {C.}~\bibnamefont {Faraco}}, \bibinfo {author} {\bibfnamefont
  {L.~S.}\ \bibnamefont {Miller}}, \ and\ \bibinfo {author} {\bibfnamefont
  {T.}~\bibnamefont {Liu}},\ }\href@noop {} {\bibfield  {journal} {\bibinfo
  {journal} {Journal of Theoretical Biology}\ }\textbf {\bibinfo {volume}
  {264}},\ \bibinfo {pages} {467} (\bibinfo {year} {2010})}\BibitemShut
  {NoStop}%
\bibitem [{\citenamefont {Manyuhina}\ \emph {et~al.}(2014)\citenamefont
  {Manyuhina}, \citenamefont {Mayett},\ and\ \citenamefont
  {Schwarz}}]{manyuhina2014}%
  \BibitemOpen
  \bibfield  {author} {\bibinfo {author} {\bibfnamefont {O.}~\bibnamefont
  {Manyuhina}}, \bibinfo {author} {\bibfnamefont {D.}~\bibnamefont {Mayett}}, \
  and\ \bibinfo {author} {\bibfnamefont {J.}~\bibnamefont {Schwarz}},\
  }\href@noop {} {\bibfield  {journal} {\bibinfo  {journal} {New Journal of
  Physics}\ }\textbf {\bibinfo {volume} {16}},\ \bibinfo {pages} {123058}
  (\bibinfo {year} {2014})}\BibitemShut {NoStop}%
\bibitem [{\citenamefont {Sudarov}\ and\ \citenamefont
  {Joyner}(2007)}]{sudarov}%
  \BibitemOpen
  \bibfield  {author} {\bibinfo {author} {\bibfnamefont {A.}~\bibnamefont
  {Sudarov}}\ and\ \bibinfo {author} {\bibfnamefont {A.~L.}\ \bibnamefont
  {Joyner}},\ }\href@noop {} {\bibfield  {journal} {\bibinfo  {journal} {Neural
  Development}\ }\textbf {\bibinfo {volume} {2}},\ \bibinfo {pages} {26}
  (\bibinfo {year} {2007})}\BibitemShut {NoStop}%
\bibitem [{\citenamefont {Halfter}\ \emph {et~al.}(2002)\citenamefont
  {Halfter}, \citenamefont {Dong}, \citenamefont {Yip}, \citenamefont
  {Willem},\ and\ \citenamefont {Mayer}}]{pial_membrane_2002}%
  \BibitemOpen
  \bibfield  {author} {\bibinfo {author} {\bibfnamefont {W.}~\bibnamefont
  {Halfter}}, \bibinfo {author} {\bibfnamefont {S.}~\bibnamefont {Dong}},
  \bibinfo {author} {\bibfnamefont {Y.-P.}\ \bibnamefont {Yip}}, \bibinfo
  {author} {\bibfnamefont {M.}~\bibnamefont {Willem}}, \ and\ \bibinfo {author}
  {\bibfnamefont {U.}~\bibnamefont {Mayer}},\ }\href@noop {} {\bibfield
  {journal} {\bibinfo  {journal} {J. Neuroscience}\ }\textbf {\bibinfo {volume}
  {22}},\ \bibinfo {pages} {6029} (\bibinfo {year} {2002})}\BibitemShut
  {NoStop}%
\bibitem [{\citenamefont {Baumann}(2014)}]{basement_membrane_2014}%
  \BibitemOpen
  \bibfield  {author} {\bibinfo {author} {\bibfnamefont {K.}~\bibnamefont
  {Baumann}},\ }\href@noop {} {\bibfield  {journal} {\bibinfo  {journal}
  {Nature Reviews Molecular Cell Biology}\ }\textbf {\bibinfo {volume} {15}},\
  \bibinfo {pages} {767} (\bibinfo {year} {2014})}\BibitemShut {NoStop}%
\bibitem [{\citenamefont {Nayfeh}\ and\ \citenamefont
  {Mook}(2008)}]{nayfeh2008nonlinear}%
  \BibitemOpen
  \bibfield  {author} {\bibinfo {author} {\bibfnamefont {A.~H.}\ \bibnamefont
  {Nayfeh}}\ and\ \bibinfo {author} {\bibfnamefont {D.~T.}\ \bibnamefont
  {Mook}},\ }\href@noop {} {\emph {\bibinfo {title} {Nonlinear Oscillations}}}\
  (\bibinfo  {publisher} {John Wiley \& Sons},\ \bibinfo {year}
  {2008})\BibitemShut {NoStop}%
\bibitem [{\citenamefont {Hayashi}(2014)}]{hayashi2014nonlinear}%
  \BibitemOpen
  \bibfield  {author} {\bibinfo {author} {\bibfnamefont {C.}~\bibnamefont
  {Hayashi}},\ }\href@noop {} {\emph {\bibinfo {title} {Nonlinear Oscillations
  in Physical Systems}}}\ (\bibinfo  {publisher} {Princeton University Press},\
  \bibinfo {year} {2014})\BibitemShut {NoStop}%
\bibitem [{\citenamefont {Brau}\ \emph {et~al.}(2011)\citenamefont {Brau},
  \citenamefont {Vandeparre}, \citenamefont {Sabbah}, \citenamefont {Poulard},
  \citenamefont {Boudaoud},\ and\ \citenamefont {Damman}}]{brau2011}%
  \BibitemOpen
  \bibfield  {author} {\bibinfo {author} {\bibfnamefont {F.}~\bibnamefont
  {Brau}}, \bibinfo {author} {\bibfnamefont {H.}~\bibnamefont {Vandeparre}},
  \bibinfo {author} {\bibfnamefont {A.}~\bibnamefont {Sabbah}}, \bibinfo
  {author} {\bibfnamefont {C.}~\bibnamefont {Poulard}}, \bibinfo {author}
  {\bibfnamefont {A.}~\bibnamefont {Boudaoud}}, \ and\ \bibinfo {author}
  {\bibfnamefont {P.}~\bibnamefont {Damman}},\ }\href@noop {} {\bibfield
  {journal} {\bibinfo  {journal} {Nature Physics}\ }\textbf {\bibinfo {volume}
  {7}},\ \bibinfo {pages} {56} (\bibinfo {year} {2011})}\BibitemShut {NoStop}%
\bibitem [{\citenamefont {Karzbrun}\ \emph {et~al.}(2018)\citenamefont
  {Karzbrun}, \citenamefont {Kshirsagar}, \citenamefont {Cohen}, \citenamefont
  {Hanna},\ and\ \citenamefont {Reiner}}]{karzbrun2018}%
  \BibitemOpen
  \bibfield  {author} {\bibinfo {author} {\bibfnamefont {E.}~\bibnamefont
  {Karzbrun}}, \bibinfo {author} {\bibfnamefont {A.}~\bibnamefont
  {Kshirsagar}}, \bibinfo {author} {\bibfnamefont {S.~R.}\ \bibnamefont
  {Cohen}}, \bibinfo {author} {\bibfnamefont {J.~H.}\ \bibnamefont {Hanna}}, \
  and\ \bibinfo {author} {\bibfnamefont {O.}~\bibnamefont {Reiner}},\
  }\href@noop {} {\bibfield  {journal} {\bibinfo  {journal} {Nature Physics}\
  }\textbf {\bibinfo {volume} {14}},\ \bibinfo {pages} {515} (\bibinfo {year}
  {2018})}\BibitemShut {NoStop}%
\bibitem [{\citenamefont {Kolb}(2018)}]{kol18}%
  \BibitemOpen
  \bibfield  {author} {\bibinfo {author} {\bibfnamefont {H.}~\bibnamefont
  {Kolb}},\ }in\ \href@noop {} {\emph {\bibinfo {booktitle} {Webvision: The
  Organization of the Retina and Visual System}}},\ \bibinfo {editor} {edited
  by\ \bibinfo {editor} {\bibfnamefont {H.}~\bibnamefont {Kolb}}, \bibinfo
  {editor} {\bibfnamefont {R.}~\bibnamefont {Nelson}}, \bibinfo {editor}
  {\bibfnamefont {E.}~\bibnamefont {Fernandez}}, \ and\ \bibinfo {editor}
  {\bibfnamefont {B.}~\bibnamefont {Jones}}}\ (\bibinfo  {publisher}
  {Webvision},\ \bibinfo {year} {2018})\ Chap.~\bibinfo {chapter}
  {1}\BibitemShut {NoStop}%
\bibitem [{\citenamefont {Riccobelli}\ and\ \citenamefont
  {Bevilacqua}(2020)}]{riccobelli}%
  \BibitemOpen
  \bibfield  {author} {\bibinfo {author} {\bibfnamefont {D.}~\bibnamefont
  {Riccobelli}}\ and\ \bibinfo {author} {\bibfnamefont {G.}~\bibnamefont
  {Bevilacqua}},\ }\href@noop {} {\bibfield  {journal} {\bibinfo  {journal}
  {Journal of the Mechanics and Physics of Solids}\ }\textbf {\bibinfo {volume}
  {134}},\ \bibinfo {pages} {103745} (\bibinfo {year} {2020})}\BibitemShut
  {NoStop}%
\bibitem [{\citenamefont {Rakic}(2009)}]{rakic2009}%
  \BibitemOpen
  \bibfield  {author} {\bibinfo {author} {\bibfnamefont {P.}~\bibnamefont
  {Rakic}},\ }\href@noop {} {\bibfield  {journal} {\bibinfo  {journal} {Nature
  Reviews Neuroscience}\ }\textbf {\bibinfo {volume} {10}},\ \bibinfo {pages}
  {724} (\bibinfo {year} {2009})}\BibitemShut {NoStop}%
\bibitem [{\citenamefont {Moln{\'a}r}\ \emph {et~al.}(2019)\citenamefont
  {Moln{\'a}r}, \citenamefont {Clowry}, \citenamefont {{\v{S}}estan},
  \citenamefont {Alzu'bi}, \citenamefont {Bakken}, \citenamefont {Hevner},
  \citenamefont {H{\"u}ppi}, \citenamefont {Kostovi{\'c}}, \citenamefont
  {Rakic}, \citenamefont {Anton} \emph {et~al.}}]{molnar2019}%
  \BibitemOpen
  \bibfield  {author} {\bibinfo {author} {\bibfnamefont {Z.}~\bibnamefont
  {Moln{\'a}r}}, \bibinfo {author} {\bibfnamefont {G.~J.}\ \bibnamefont
  {Clowry}}, \bibinfo {author} {\bibfnamefont {N.}~\bibnamefont
  {{\v{S}}estan}}, \bibinfo {author} {\bibfnamefont {A.}~\bibnamefont
  {Alzu'bi}}, \bibinfo {author} {\bibfnamefont {T.}~\bibnamefont {Bakken}},
  \bibinfo {author} {\bibfnamefont {R.~F.}\ \bibnamefont {Hevner}}, \bibinfo
  {author} {\bibfnamefont {P.~S.}\ \bibnamefont {H{\"u}ppi}}, \bibinfo {author}
  {\bibfnamefont {I.}~\bibnamefont {Kostovi{\'c}}}, \bibinfo {author}
  {\bibfnamefont {P.}~\bibnamefont {Rakic}}, \bibinfo {author} {\bibfnamefont
  {E.}~\bibnamefont {Anton}},  \emph {et~al.},\ }\href@noop {} {\bibfield
  {journal} {\bibinfo  {journal} {Journal of Anatomy}\ }\textbf {\bibinfo
  {volume} {235}},\ \bibinfo {pages} {432} (\bibinfo {year}
  {2019})}\BibitemShut {NoStop}%
\bibitem [{\citenamefont {Van~Essen}(1997)}]{van1997}%
  \BibitemOpen
  \bibfield  {author} {\bibinfo {author} {\bibfnamefont {D.~C.}\ \bibnamefont
  {Van~Essen}},\ }\href@noop {} {\bibfield  {journal} {\bibinfo  {journal}
  {Nature}\ }\textbf {\bibinfo {volume} {385}},\ \bibinfo {pages} {313}
  (\bibinfo {year} {1997})}\BibitemShut {NoStop}%
\bibitem [{\citenamefont {Van~Essen}(2020)}]{van2020}%
  \BibitemOpen
  \bibfield  {author} {\bibinfo {author} {\bibfnamefont {D.~C.}\ \bibnamefont
  {Van~Essen}},\ }\href@noop {} {\bibfield  {journal} {\bibinfo  {journal}
  {Proceedings of the National Academy of Sciences}\ }\textbf {\bibinfo
  {volume} {117}},\ \bibinfo {pages} {32868} (\bibinfo {year}
  {2020})}\BibitemShut {NoStop}%
\end{thebibliography}%

\section{Appendix}
\subsection{Uncoupling general Euler-Lagrange \\equations for $r,t$}\label{general uncoupled}
The uncoupled nonlinear differential equation in $r$ for the case of having $k_r:=k_r(r)$ is shown in Eq. \ref{general_kr_EL}. The form of such an equation is, however, dependent on the choice of nonlinear coefficients. For general coupled Euler-Lagrange equations, we have
\begin{equation}\label{f}
f\va=0,
\end{equation}
\begin{equation}
t\ppr=g\va.
\end{equation} 
Here the prime superscript indicates a derivative with respect to $\theta$. The derivative of Eq. \ref{f} with respect to $\theta$ gives,
\begin{equation}\label{fpr}
\frac{\partial f\va}{\partial r}r\pr+\frac{\partial f\va}{\partial t}t\pr=0.
\end{equation}
Another derivative yields,
\begin{equation}
\begin{split}
\frac{\partial f\va}{\partial r}r\ppr+\frac{\partial f\va}{\partial t}t\ppr+\frac{\partial ^2 f\va}{\partial r^2} r^{\prime 2}&+2\;\frac{\partial^2 f\va}{\partial r \partial t}r\pr t\pr\\
&+\frac{\partial^2 f\va}{\partial t^2}t^{\prime 2}=0.
\end{split}
\end{equation}
Eq. \ref{f},\ref{fpr} can be solved to find $t:=t(r),\;t\pr=t\pr(r,r\pr)$. The uncoupled differential equation in $r$ would then be,
\begin{equation}
\begin{split}
&\frac{\partial f\var}{\partial r}r\ppr+\frac{\partial f\var}{\partial t}t\ppr(r)+\frac{\partial ^2 f\var}{\partial r^2} r^{\prime 2}\\
&+2\;\frac{\partial^2 f\var}{\partial r \partial t}r\pr t\pr(r)+\frac{\partial^2 f\var}{\partial t^2}t^{\prime 2}(r)=0.
\end{split}
\end{equation}  
The last two terms for Eq. \ref{general_kr_EL} is absent and this particular case of nonlinear differential equation is of the form,
\begin{equation}
\frac{\partial f\var}{\partial r}r\ppr+\frac{\partial ^2 f\var}{\partial r^2} r^{\prime 2}+\frac{\partial f\var}{\partial t}g(r,t(r))=0,
\end{equation}
where the implicit dependence - $t(r)$ is to be taken into account \textit{after} evaluating the partial derivatives in $t,r$. This renders an uncoupled differential equation of motion for $r$.

\subsection{Higher order corrections to the Simple Harmonic Oscillator}\label{cubic_quartic_HO}
With no loss of generality, we can study the simple harmonic oscillator with cubic and quartic corrections to the potential energy as,

\begin{figure}[thb]
\includegraphics[width=0.48\textwidth]{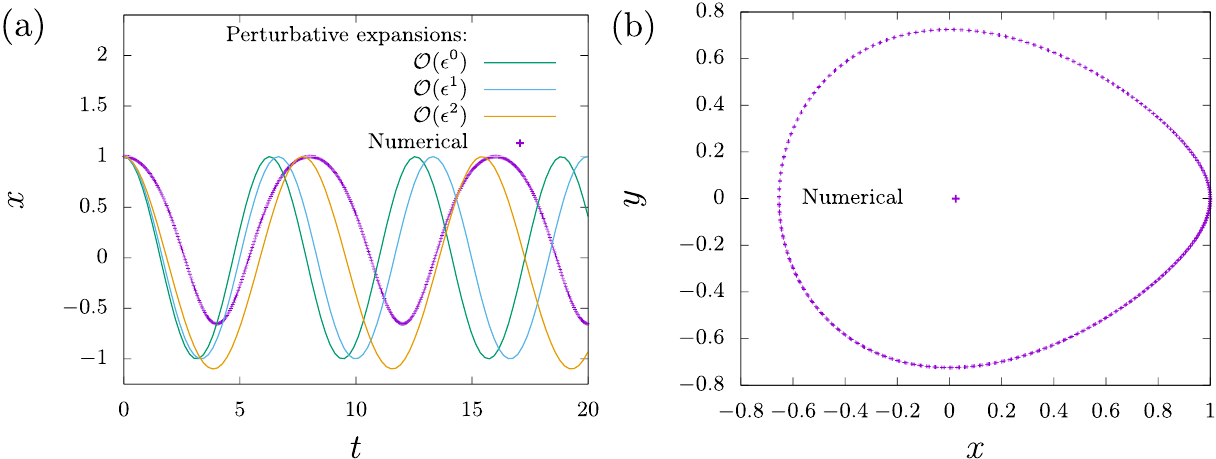}
\caption{\textit{Lindstedt-Poincar\'e perturbation method for the cubic and quartic energy corrections to the SHO.} $\mathcal{O}(\epsilon^2)$ curve follows the leading edge of the numerical solution but doesn't capture the offset in troughs and the relative difference in widths of crests and troughs. Here $\epsilon=0.6,\Gamma^{\prime}=0.25$.}\label{lp_fig}
\end{figure}
\begin{equation}\label{non_dimensional_anharmonic}
\frac{d^2x}{dt^2}=-x-x^2-\Gamma^{\prime} x^3,
\end{equation}
a single parameter equation where all variables are nondimensionalized. To evolve a perturbative scheme, we treat the last two terms of the above equation as a perturbation to the simple harmonic oscillator. To that effect, with $\epsilon>0$ we have,

\begin{equation}\label{epsilon}
\frac{d^2x}{dt^2}=-x-\epsilon\;(x^2-\Gamma^{\prime} x^3).
\end{equation}

The Lindstedt-Poincar\'e method \cite{jordan2007} introduces angular frequency $\omega$ by change of variable $\tau=\omega t$. This allows for coupling between the frequency and amplitude. Eq. \ref{epsilon} is now written as,

\begin{equation}\label{omega}
\omega^2\frac{d^2 x}{d\tau^2}+x+\epsilon x^2-\epsilon \Gamma^\prime x^3=0.
\end{equation}

\noindent The series expansions for $x$ and $\omega$ are,

\begin{figure*}[tbh]
\includegraphics[width=1\textwidth]{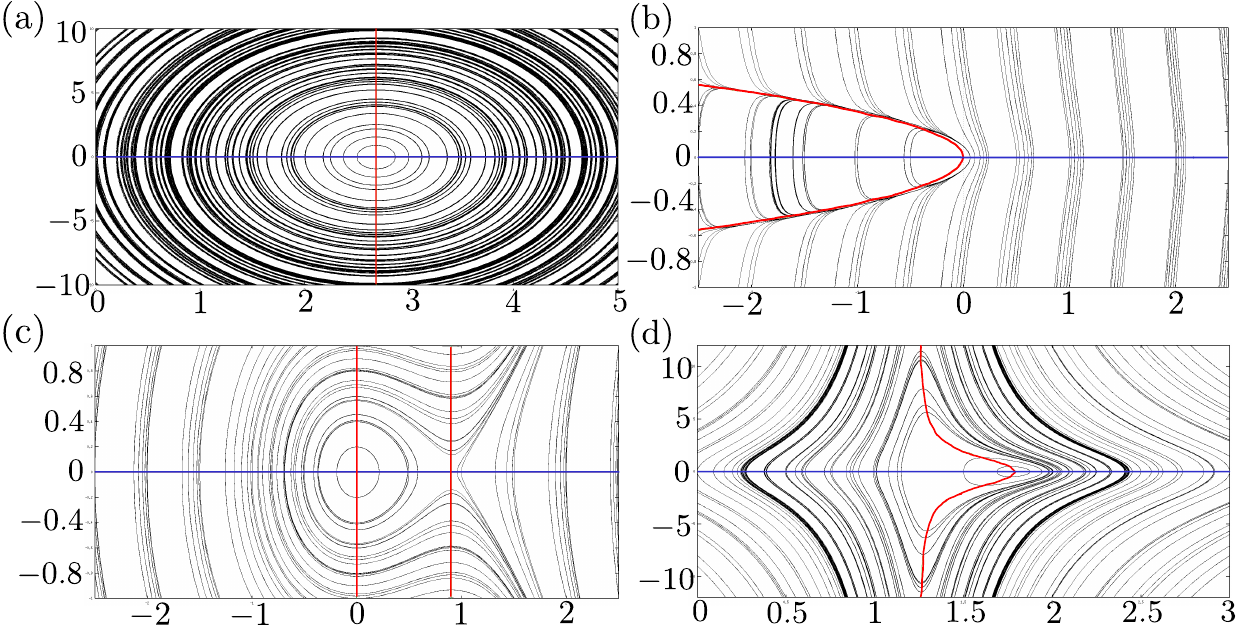}
\caption{\textit{Phase portraits of BWBM and ADO systems}. Blue lines are position nullclines: $dr/dt=0$ and red lines are velocity nullclines: $d^2r/dt^2=0$. The tangent to the flow lines are vertical when they pass the position nullcline and horizontal when they pass the velocity nullcline. a)Linear BWBM b) ADO c) Cubic and quartic energy corrections to the SHO d) BWBM with nonharmonic springs. In (b,c,d) left-right asymmetric orbits where there is a steeper fall on the left translates to sharper troughs in position-time space. We see left-right asymmetric orbits with or without bent nullclines (see b,d \& c).}\label{phase portraits panel}
\end{figure*}
\begin{equation}\label{series expansion}
\begin{split}
x(t)&=x_0(t)+\epsilon x_1(t)+\epsilon^2 x_2(t)+ ...\\
\omega&=1+\epsilon \omega_1+\epsilon^2 \omega_2 +...
\end{split}
\end{equation}

\noindent The boundary conditions are,
\begin{equation}\label{bc}
\begin{split}
x_0(0)&=A_0,\dot{x}_0(0)=0,\\
x_i(0)&=0,\dot{x}_i(0)=0~~i>0.
\end{split}
\end{equation}
\begin{large}\underline{$\mathcal{O}(\epsilon^0)$}\end{large}
\\We substitute the series expansions of Eq. \ref{series expansion} in Eq. \ref{omega}. All the terms at every given order of $\epsilon$ must add up to zero for Eq. \ref{omega} to hold. The response equation at the zeroth order of $\epsilon$ then is,
\begin{equation}
\frac{d^2x_0}{d\tau^2}+x_0=0,
\end{equation}
which is just the equation of motion of the simple harmonic oscillator. The zeroth order response that obeys the boundary conditions \ref{bc} is,
\begin{equation}\label{epsilon0}
x_0(\tau)=A_0\co(\tau).
\end{equation}
\begin{large}\underline{$\mathcal{O}(\epsilon^1)$}\end{large}
\\At the first order of $\epsilon$ we have,
\begin{equation}\label{epsilon1}
\frac{d^2x_1}{d\tau^2}+x_1=x_0^2+\Gamma^{\prime} x_0^3-2\omega_1 \frac{d^2x_0}{d\tau^2}.
\end{equation}

\noindent Solutions of lower order response equations are used to find solutions of higher order response equations. $\omega_1$ is fixed by the demand that the solution to Eq. \ref{epsilon1} be periodic. For this, upon substituting the zeroth order solution (see Eq. \ref{epsilon0}) in Eq. \ref{epsilon1} we set the coefficient of cos$(\tau)$ in the resulting RHS of Eq. \ref{epsilon1} to vanish. We obtain 

\begin{equation}
\omega_1=\frac{-3\Gamma^{\prime} a_0^2}{8}.
\end{equation}

\noindent The amplitude-frequency dependence can be seen here. The first order correction to $x$ which obeys the boundary conditions is,
\begin{equation}
\begin{split}
x_1(\tau)=A_0^2\left(\frac{\Gamma^{\prime} A_0}{32}-\frac{1}{3}\co(\tau)\right)+\frac{A_0^2}{2}&-\frac{A_0^2}{6}\co(2\tau)\\
&-\frac{\Gamma^{\prime} A_0^3}{32}\co(3\tau). 
\end{split}
\end{equation}
\begin{large}\underline{$\mathcal{O}(\epsilon^2)$}\end{large}
\\At the second order of $\epsilon$, we have,
\begin{equation}
\frac{d^2x_2}{d\tau^2}+x_2=3\Gamma^{\prime} x_0^2x_1+2x_0x_1-(\omega_1^2+2\omega_2)\frac{d^2x_0}{d\tau^2}-2\omega_1\frac{d^2x_1}{d\tau^2}.
\end{equation}
Carrying out the same procedure at this order, we have,
\begin{equation}
\begin{split}
\omega_2=\frac{1}{384} \bigg(-9 a_0^4 \Gamma^{\prime 2}
    +144 a_0^3 \Gamma^{\prime} &-12
    a_0^2 \Gamma^{\prime}  \omega_1-160 a_0^2\\
&+128 a_0
    \omega_1-192 \omega_1^2\bigg), 
\end{split}
\end{equation}
and
\begin{equation}
\begin{split}
x_2(\tau)=&-\frac{1}{384} a_0^2
    [-64
    \co (2 \tau ) (3 a_0^2
    \Gamma^{\prime} -2 a_0-8 \omega_1)\\
    &+a_0 (\co (3 \tau )
    (9 a_0^2 \Gamma^{\prime 2} +96
    a_0 \Gamma^{\prime} +216 \Gamma^{\prime} 
    \omega_1+64)\\
    &+60 a_0 \Gamma^{\prime}  \co (4 \tau
    )+9
    a_0^2 \Gamma^{\prime 2} \co (5 \tau
    )
    -252 a_0 \Gamma^{\prime}\\
    & +128)].
\end{split}
\end{equation}
The perturbative expansions are compared with numerical solution in Fig. \ref{lp_fig}.

\vspace{20pt}
\subsection{Phase portraits}
Closed orbits in phase-space translate to periodic motion in the position-time space. Plotting orbits in phase space for different initial conditions builds the phase portrait of the system. The $x\; (dx/dt)$ nullcline are the locus of points where $dx/dt=0$ ($d^2x/dt^2=0$). The nullclines are generally used to divide the phase-portrait into regions where the tangent of the orbit points in the same general direction (NW, NE, SE or SW). The orbit has a vertical (horizontal) tangent when it passes through the $x$ ($dx/d\theta$) nullcline. The intersection of the nullclines gives the equilibrium point where both the `velocity' and `acceleration' of the system is zero. The time-reversal symmetry of the governing differential equations (see Eq. \ref{general_kr_EL}, \ref{ADO}, \ref{nondim_tanh_circ}) guarantees closed orbits at points close enough to the equilibrium point. This is verified in the phase-portraits in Fig. \ref{phase portraits panel}.

\end{document}